\documentclass[12pt]{article}

\usepackage{amsmath,amsthm, amsfonts, amssymb, amsxtra, amsopn}
\usepackage{pgfplots}
\usepgfplotslibrary{colorbrewer}
\pgfplotsset{compat = 1.15, 
			 cycle list/Set1-3} 
\usetikzlibrary{pgfplots.statistics, pgfplots.colorbrewer} 
\usepackage{pgfplotstable}
\usepackage{graphicx,grffile}
\usepackage{multirow}
\usepackage{longtable} 
\usepackage{booktabs}
\usepackage{tcolorbox}
\usepackage{algorithm} 
\usepackage[noend]{algpseudocode} 
\usepackage{listings}
\usepackage{cmap}
\usepackage{colortbl}
\usepackage{adjustbox}
\usepackage{epsfig}
\usepackage{bm}

\usepackage[tableposition=top,font=small,skip=5pt]{caption}
\usepackage{subcaption}
\usepackage{makecell}

\usepackage[explicit]{titlesec}

\usetikzlibrary{patterns}

\def\bbb#1{{\color{blue}#1}}
\def\com#1{\bbb{\texttt{/\kern-1.5pt /} #1}}

\def\tau{\mathcal{T}}

\PassOptionsToPackage{hyphens}{url}
\PassOptionsToPackage{table}{xcolor}

\usepackage{hyperref}
\hypersetup{colorlinks=true,linkcolor=black,citecolor=black,urlcolor=blue,filecolor=black}
\hypersetup{pdfpagemode=UseNone,pdfstartview=}

\definecolor{darkgreen}{rgb}{0.125,0.5,0.169}

\usepackage[shortlabels]{enumitem}
\setlist[itemize]{noitemsep, topsep=0pt}

\advance\oddsidemargin by -0.45in
\advance\textwidth by 0.9in

\advance\topmargin by -0.5in
\advance\textheight by 1.0in

\long\def\symbolfootnotetext[#1]#2{\begingroup%
  \def\thefootnote{\fnsymbol{footnote}}\footnotetext[#1]{#2}\endgroup}

\newcommand\dunderline[3][-1pt]{{%
      \sbox0{#3}%
      \ooalign{\copy0\cr\rule[\dimexpr#1-#2\relax]{\wd0}{#2}}}}
\def\uuu{\kern-1pt\dunderline{0.75pt}{\phantom{M}}}

\def\grayscale{\textrm{Grayscale}}
\def\entropyHilbert{\textrm{Entropy Hilbert}}
\def\byteclassHilbert{\textrm{Byteclass Hilbert}}
\def\hit{\textrm{HIT}}
\def\bigramCartesian{\textrm{Bigram Cartesian}}
\def\bigramPolar{\textrm{Bigram Polar}}
\def\spiral{\textrm{Spiral}}
\def\byteclass{\textrm{Byteclass}}

\DeclareMathOperator*{\argmax}{arg\,max}

\def\zz{\phantom{0}}

\title{Image-Based Techniques and Ensemble Soft Voting for Malware Classification}

\author{Sushant Rakesh Lokhande\footnotemark[1]\ \ \ 
Fabio Di Troia\footnotemark[1]\ \ \ 
Martin  Jure\v{c}ek\footnotemark[2]\ \ \
Mark Stamp\footnotemark[1]\,\,\footnotemark[3]} 

\begin{document}

\symbolfootnotetext[1]{Department of Computer Science, San Jose State University}
\symbolfootnotetext[2]{Faculty of Information Technology, Czech Technical University in Prague}
\symbolfootnotetext[3]{mark.stamp$@$sjsu.edu}

\maketitle

\abstract
In this chapter, we investigate image-based malware family classification 
using an ensemble learning framework and a soft voting strategy. We consider malware
binaries that have been converted into images using eight distinct conversion
strategies. Three complementary feature extraction tracks are applied
to these images: handcrafted descriptors combining Histogram of
Oriented Gradients (HOG) and Haralick texture features along with~38 statistical features;
dense embeddings obtained from three pretrained neural networks (VGG16,
ResNet50, and ViT-B/16), where each pretrained model is used as a frozen feature extractor with its classification head removed; 
and~512-dimensional embeddings derived from a custom Convolutional Neural Network (CNN)
trained directly on the malware images. Each of the three feature extraction techniques 
is evaluated with machine learning classifiers across all eight image conversion types. 
The best individual results are~77.8\%\ accuracy for the handcrafted features,
73.8\%\ for the pretrained neural network track, and~74.8\%\ for the custom CNN track. 
Then we consider various soft voting ensemble strategies, and we
find that the best-performing soft voting pool---consisting of fifteen voters selected on a dedicated
validation split---achieves~80.2\%\ accuracy across the~17 malware families
under consideration, a statistically significant improvement of~2.4 percentage
points over the best individual model. A quantitative diversity analysis confirms
that the different feature representations are complementary, with the
handcrafted descriptors being the strongest contributors.

\bigskip


\section{Introduction}\label{sec:intro}

Malware continues to be a significant threat to computer systems and networks
worldwide. Traditional detection methods rely on identifying known byte patterns or
behavioral signatures in executable files. 
These approaches struggle when malware authors deliberately obfuscate
the binary structure of a file to evade detection. 
This has motivated researchers to explore
representations of malware that capture higher-level structural properties and are
more robust to obfuscation techniques.

One approach that has gained considerable prominence is treating malware binaries as
images. By mapping the raw bytes 
or other features extracted from an executable file to pixel intensities arranged
in a two-dimensional grid, malware samples can be analyzed using advanced computer
vision and machine learning techniques. 
Nataraj et al.~\cite{Nataraj_2011} showed that the bytes 
from malware families produce
visually distinguishable patterns under image representation, and that these
patterns can support accurate classification. Many additional studies have
explored different image representations and feature 
extraction strategies~\cite{Gibert_2019,Vasan_2020}.

In this chapter, we consider a feature engineering and ensemble classification
system for malware family identification, building on the dataset and image conversion
pipeline introduced by Agrawal et al.~\cite{Agrawal_2025}. Our work proceeds across
three feature extraction tracks. The first of these uses handcrafted image descriptors,
specifically Histogram of Oriented Gradients (HOG)~\cite{Dalal_2005}, Haralick
texture features~\cite{Haralick_1973}, and a set of~38 statistical features we refer to 
as ADV38 (Advanced~38)~\cite{Chavda_2020}, combined and reduced 
using a two step Random Forest based feature
selection process~\cite{Breiman_2001}. The second track uses pretrained convolutional neural
networks as fixed feature extractors---specifically, we apply VGG16~\cite{Simonyan_2015},
ResNet50~\cite{He_2016}, and ViT~\cite{Dosovitskiy_2021} to produce dense embedding
vectors from malware images. Our third track trains a custom CNN architecture directly on
malware images to produce embeddings. Finally, the top-performing
models from all three tracks are combined into an
ensemble via soft voting~\cite{Dietterich_2000}, where each model contributes a probability
distribution over the~17 malware families under consideration,
and the distributions are averaged to produce a final prediction. In addition,
we quantify the diversity of the voting pools and assess the statistical
significance of the main comparisons using bootstrap confidence intervals and
McNemar's test~\cite{McNemar}.

The malware image dataset used throughout this work was developed in~\cite{Agrawal_2025}.
This dataset consists of~17 malware families with~1,000
samples per family. For each of these~17,000 samples, eight images have been generated,
based on distinct features. All evaluation is performed on a held-out test set 
of~1,700 samples that is never used during any stage of model training or selection.


Image-based malware classification considers the overall visual structure of a binary rather than 
looking for specific byte sequences. Handcrafted image descriptors are interpretable 
and compact, but they depend on manually specified feature
designs. Pretrained CNN embeddings capture hierarchical visual patterns but
these models were trained on natural images that differ significantly from malware
visualizations. A custom CNN trained on malware images can potentially learn
representations specific to the malware domain. Each of these three tracks 
has different strengths, and hence there may be scope for better classification 
when combined via ensembles.

In this chapter, we explore all three tracks under the same experimental conditions
and evaluate whether ensemble techniques 
can outperform the individual tracks. The results provide a systematic comparison of
handcrafted and learned feature representations for malware image classification,
and demonstrate the value of combining diverse features.


As mentioned above, we apply three distinct feature
extraction pipelines to the malware images in our dataset. 
The primary research question 
considered in this chapter is the following:
Does combining feature representations from handcrafted descriptors, pretrained neural network
embeddings, and custom CNN embeddings improve malware family classification accuracy
beyond that achieved by the individual feature types?

To resolve this research question, we address each of the following subsidiary questions.
\begin{itemize}
    \item How well do embeddings extracted from pretrained neural network models 
    perform for malware image classification when paired with classical machine learning techniques?\\[-1.75ex]
    \item Does a custom CNN trained on malware images produce more discriminative
    embeddings than pretrained models designed for natural images?\\[-1.75ex]
    \item Which image conversion types and feature extraction techniques produce the strongest
    individual classifiers?\\[-1.75ex]
    \item Can a soft voting ensemble using the top-performing models 
    improve the results beyond the best individual model?\\[-1.75ex]
    \item Which malware families are the most difficult to classify across the 
    various models considered?\\[-1.75ex]
\end{itemize}


The following steps are performed to address the main research question, and the subsidiary
questions listed above.
\begin{enumerate}
    \item We extract and evaluate handcrafted feature representations (HOG, Haralick,
    ADV38) across eight malware image conversion types. Then we apply a two-stage Random
    Forest based feature reduction to identify relatively compact discriminative subsets.
    \item We extract and save dense embedding vectors from pretrained CNN and transformer
    architectures (VGG16, ResNet50, and ViT-B/16) applied to malware images across all eight
    conversion types. We then evaluate classical machine learning classifiers trained on
    these embeddings.
    \item We design and train a custom CNN architecture on malware images, extract
    embeddings from its penultimate layer, and evaluate the same set
    of classical classifiers on these embeddings.
    \item We construct a soft voting ensemble using the top-performing configurations
    from each feature extraction technique, and evaluate whether the ensemble improves classification
    accuracy on the test set.
    \item We analyze per-class performance and the most commonly confused malware family pairs to 
    characterize where our best ensemble model succeeds and where errors remain.
\end{enumerate}

The remainder of this chapter is organized as follows.
Section~\ref{sec:background} provides background on image-based malware
classification, the eight conversion techniques, the feature descriptors,
and the machine learning models used throughout this work.
Section~\ref{sec:implementation} describes the dataset, the three feature
extraction tracks, and the ensemble construction.
Section~\ref{sec:results} presents our experimental results for each
track and for the soft voting ensemble, including per-class performance
and confusion analysis. Section~\ref{sec:conclusion} concludes the chapter
and discusses directions for future work.

\section{Background}\label{sec:background}

In this section, we first discuss the most relevant prior research. Then we introduce each of the
eight binary-to-image conversion techniques under consideration, and we discuss each of
our three feature extraction techniques (i.e., handcrafted features, features extracted from pretrained 
neural networks, and features from a custom CNN). We also outline our ensemble strategy, and we conclude
this section with an overview of the machine learning techniques that we use for classification.

\subsection{Image-Based Malware Classification}\label{sec:bg_malware}

Traditionally, malware detection relied on signature-based methods, which identify known
byte sequences or behavioral patterns in executable files. While these methods can be
effective against many types of previously-observed threats, 
they are vulnerable to advanced malware
techniques, including obfuscation and
polymorphism, where malware authors modify the binary
structure of a file to evade detection while preserving the malicious behavior.
This limitation has motivated the development of alternative representations that
capture higher-level structural properties of malware.

The idea of visualizing malware binaries as images was introduced by Nataraj et
al.~\cite{Nataraj_2011}, who observed that when the bytes of an executable file are
considered as pixel intensities in a two-dimensional grid, malware samples from the same
family tend to produce visually similar images. 
These authors demonstrated that standard image texture features---specifically
Gabor filters applied to grayscale malware images---could achieve strong classification
accuracy without requiring disassembly or execution.

Gibert et al.~\cite{Gibert_2019} propose a deep learning approach that groups
malware into families based on discriminant patterns extracted from their visual
representations, demonstrating that convolutional neural networks can learn these
patterns directly from raw images. Vasan et al.~\cite{Vasan_2020} introduced IMCFN,
a fine-tuned CNN architecture applied to color malware images that achieves strong
accuracy on benchmark datasets by leveraging transfer learning via the ImageNet architecture. 
These papers established that both handcrafted and learned features can be effective for
malware image classification, and motivated the comparisons considered in the research 
presented in this chapter.

Agrawal et al.~\cite{Agrawal_2025} provide a systematic comparison of eight
malware-to-image conversion strategies across a range of learning models. Their
results showed that several conversions perform similarly despite producing visually
distinct images, suggesting that the use of advanced visualization-based classification techniques
might matter more than the specific binary-to-image conversion method. We use this
same dataset for the research reported in this chapter.

\subsection{Malware-to-Image Conversion Techniques}\label{sec:conversions}

Every malware binary considered in this work is transformed into a~$224 \times 224$ image
based on each of eight conversion strategies. These strategies differ in how they map
the bytes of an executable file into a two-dimensional pixel
representation. Each conversion is designed to highlight a different structural property 
of the underlying binary, providing a distinct view of the same malware sample.
Next, we provide a brief description of each of these image conversion 
techniques---for more details, see~\cite{Agrawal_2025}.

\subsubsection{\grayscale}\label{sec:grayscale}

\grayscale\ is the simplest conversion strategy, 
as it maps each byte directly to a pixel intensity value
between~0 and~255. Each binary is represented using the
first~$224 \times 224 = 50{,}176$ bytes; files larger than this are trimmed, while
smaller files are zero-padded. These bytes are then arranged row by row into a
$224 \times 224$ two-dimensional grid. Despite its simplicity, the \grayscale\ conversion generally
produces competitive classification results across a range of models. 

\subsubsection{\byteclass}\label{sec:byteclass}

Rather than mapping raw byte values directly to pixel intensities, the 
\byteclass\ conversion categorizes each byte into one of several ASCII classes (e.g., printable
characters, control characters, and high-byte values) and maps each category to a
distinct intensity value in the green channel of an RGB image. This approach reveals
the structural organization of different byte types within the binary and can expose
patterns related to the presence of encoded strings or executable code regions, for example.

\subsubsection{\byteclassHilbert}\label{sec:hilbert}

The \byteclassHilbert\ conversion applies the same categorization 
as in the \byteclass\ conversion,
but arranges the bytes along a Hilbert space-filling curve rather than in a linear
scan. The Hilbert curve is a continuous fractal curve that visits every point in a
two-dimensional grid while preserving spatial locality. Thus, bytes that are close
to each other in the sequence are also placed close to each other in the image.
This property can reveal structural patterns that are obscured by a row-by-row
arrangement.

\subsubsection{\entropyHilbert}\label{sec:entropy}

The \entropyHilbert\ conversion computes Shannon entropy over a sliding window of bytes as it
moves through the binary. The resulting entropy values, which encode the local
randomness or uniformity of bytes, are mapped to the red and blue channels
of an RGB image using a Hilbert traversal. Regions of high entropy typically
correspond to encrypted or compressed content, while regions of low entropy 
are indicative of repetitive data, such as padding or fixed headers.

\subsubsection{\hit}\label{sec:hit}

The Hybrid Image Transformation (\hit) 
conversion combines the entropy-based red and blue channels from the \entropyHilbert\ 
conversion with the \byteclass\ green channel. This hybrid approach blends structural
and statistical byte properties into a single image, potentially providing a richer representation
than either the \entropyHilbert\ or \byteclass\ conversions alone. 

\subsubsection{\bigramCartesian}\label{sec:bigram_cart}

The \bigramCartesian\ conversion treats consecutive byte pairs as two-dimensional
coordinates. Each pair of bytes is then interpreted as an~$(x, y)$ coordinate
in a~$256 \times 256$ grid, and the pixel intensity at that coordinate is incremented
each time the bigram appears in the binary. The resulting image represents the
frequency distribution of byte pair co-occurrences and captures statistical
dependencies between adjacent bytes.

\subsubsection{\bigramPolar}\label{sec:bigram_polar}

The \bigramPolar\ conversion applies a similar approach to \bigramCartesian, but
maps byte pairs to polar coordinates rather than Cartesian coordinates. The first byte
of each pair defines the radius and the second defines the angle---pixel intensities
are again incremented based on the bigram frequency. 

\subsubsection{\spiral}\label{sec:spiral}

The \spiral\ conversion maps normalized byte histogram values, reordered by feature
importance, into a spiral pattern within the image. Unlike the other seven conversions,
which operate on raw byte values or byte pairs, the \spiral\  conversion is derived from
statistical properties of the byte histogram.

\subsection{Handcrafted Feature Descriptors}\label{sec:handcrafted}

The first of our feature extraction tracks uses three types of handcrafted image descriptors,
each capturing different visual properties of the malware images. These descriptors
are extracted independently and concatenated into a single fused
feature vector of size~6,135
($6{,}084$ from HOG plus~$13$ from Haralick plus $38$ from ADV38),
which is then reduced using a two-stage Random Forest based selection
procedure, as described in Section~\ref{sec:implementation}.

\subsubsection{Histogram of Oriented Gradients}\label{sec:hog}

The Histogram of Oriented Gradients (HOG) descriptor was introduced by Dalal and
Triggs~\cite{Dalal_2005} for pedestrian detection, and has since become widely used
in image classification tasks. HOG divides an image into small spatial cells, computes
the gradient magnitude and orientation at each pixel within a cell, and accumulates
these into a histogram of gradient orientations. Adjacent cells are grouped into
blocks, and the histograms within each block are normalized together to reduce the
effect of illumination changes. The resulting descriptor captures local edge
and shape structure across the image.

\subsubsection{Haralick Texture Features}\label{sec:haralick}

Haralick features were introduced by Haralick, Shanmugam, and
Dinstein~\cite{Haralick_1973} and are derived from the gray-level co-occurrence
matrix (GLCM) of an image. The GLCM records how often pairs of pixel intensities
occur at a specified spatial relationship, and Haralick features summarize the
statistical properties of this matrix. The~13 Haralick features computed in this work include
contrast, correlation, energy, homogeneity, entropy, and several related measures.
These descriptors capture global texture properties that complement the local
gradient information provided by HOG. All images are resized to~$224 \times 224$ and
converted to grayscale before Haralick extraction.

\subsubsection{ADV38 Handcrafted Features}\label{sec:adv38}

Our ADV38 feature set consists of~38 handcrafted image descriptors which were used
for image spam detection by Chavda et al.~\cite{Chavda_2020}. These
features are designed to capture general-purpose image statistics that are not
domain-specific, making them suitable for adaptation to malware images. These~38
features are organized into the following five categories.

\begin{description}
\item[\textbf{Metadata features}] include height, width, aspect ratio, compression ratio,
file size, and image area. These capture basic structural properties of the image.
\item[\textbf{Color features}] include the entropy of RGB and HSV histograms, along with
the mean, variance, skewness, and kurtosis of each color channel. These describe
the underlying intensity distribution, which can differ substantially between malware
families depending on the byte composition of their binaries.
\item[\textbf{Texture features}] consist of the Local Binary Pattern (LBP) entropy, which
summarizes the spatial relationship between neighboring pixels and quantifies the
consistency of local texture patterns.
\item[\textbf{Shape features}] include the entropy of the HOG descriptor, the total number
of detected edges, and the average edge length. These describe the geometric
structure and edge density of the image at a global level.
\item[\textbf{Noise features}] include the entropy of image noise and the signal-to-noise
ratio (SNR), capturing the degree of randomness and uniformity within the image.
\end{description}

\subsection{Pretrained Neural Network Feature Extraction}\label{sec:pretrained}

Our second feature extraction track uses pretrained neural networks
as fixed feature extractors. Rather than training a classifier end-to-end, each
network is loaded with weights pretrained on the ImageNet dataset and used in inference mode
with the final classification layer removed. Malware images are passed through the
network, and the output of the penultimate layer is saved as a dense embedding vector.
This track is motivated by the work of Yosinski et al.~\cite{Yosinski_2014},
which shows that features learned by deep networks on large image datasets tend to be
general in the lower layers and task-specific in the upper layers, and that
transferring these features to new tasks often improves performance over training
from scratch---particularly when the target dataset is relatively small.

Next, we discuss each of the three pretrained architectures used in this research.
Specifically, we consider VGG16, ResNet50, and a ViT model.

\subsubsection{VGG16}\label{sec:vgg16}

VGG16, introduced by Simonyan and Zisserman~\cite{Simonyan_2015}, is a 16-layer
convolutional network that exclusively uses small~$3\times3$ convolution filters
throughout its architecture. Its depth and uniform filter size allow it to learn
rich hierarchical representations. When used as a feature extractor in this work, the
output of the penultimate fully connected layer produces a~4,096-dimensional
embedding per image.

\subsubsection{ResNet50}\label{sec:resnet}

ResNet50, introduced by He et al.~\cite{He_2016}, is a 50-layer residual network
that addresses the degradation problem observed in very deep networks by introducing
skip connections---also called residual connections---that allow gradients to flow
directly through the network during training. These connections make it possible to
train substantially deeper networks without a loss in accuracy. ResNet50 produces 
a~2,048-dimensional embedding from its global average pooling layer. 

\subsubsection{Vision Transformer (ViT)}\label{sec:vit}

The Vision Transformer (ViT), introduced by Dosovitskiy et al.~\cite{Dosovitskiy_2021}
apply a transformer self-attention mechanism to image classification by dividing
an image into fixed-size patches. Each patch is then treated as a token, and the
sequence of tokens is processed with a standard transformer encoder. 
Unlike convolutional networks, ViT does not rely on the 
local spatial biases that are built into CNN-based architectures, and
instead learns spatial relationships through attention. When pretrained on large
datasets, ViT achieves strong results comparable to state-of-the-art convolutional
networks. In this work, the ViT-B/16 variant is used (henceforth, referred to as simply ViT), 
which divides images into~$16 \times 16$ pixel patches and produces a~768-dimensional 
embedding. 

\subsection{Custom CNN Feature Extraction}\label{sec:custom_cnn}

Our third feature extraction track uses a custom CNN architecture designed and
trained directly on malware images. Following the approach of Gibert et al.~\cite{Gibert_2019} 
and Vasan et al.~\cite{Vasan_2020}, the resulting custom CNN is not used as a standalone end-to-end
classifier, but instead as a feature extractor. After training, the network weights are frozen
and the output of the penultimate layer---a 512-dimensional embedding vector---is
extracted for each image. Classical machine learning classifiers are then trained on
these embeddings. This two-stage approach separates the representation learning from
the classification step, allowing a more direct comparison with the pretrained neural network
embeddings and the handcrafted features under the same experimental conditions.

\subsection{Ensembles}\label{sec:ensemble}

Ensemble methods combine the predictions of multiple independently trained models to
produce a final output that can be more reliable than any individual model. Dietterich~\cite{Dietterich_2000}
provides a thorough analysis of why ensembles can outperform individual classifiers,
identifying the following three main conditions under which this occurs.
\begin{enumerate}
\item The individual models are accurate enough to perform above chance.
\item The individual models are sufficiently diverse in the errors they make.
\item There are enough models to allow errors to be overridden 
by correct predictions from other members of the ensemble.
\end{enumerate}

\subsubsection{Soft Voting}\label{sec:soft_voting}

Soft voting is one of the simplest and most widely used ensemble strategies. Rather
than having each model predict a single class label and taking a majority vote (i.e., hard voting), 
each model outputs a probability distribution over all classes, and these distributions
are averaged element-wise across all models. The class with the highest resulting
probability is taken as the final prediction. This approach is more informative than
hard voting because it accounts for the confidence of each model's prediction. A
model that is highly confident about a prediction contributes more to the averaged
distribution than a model that is uncertain, even if both predict the same 
class~\cite{Dietterich_2000}.

\subsection{Classical Machine Learning Models}\label{sec:classifiers}

In this research, classical machine learning models are used as
classifiers on top of the extracted feature representations. Specifically,
we consider the following classic machine learning models.

\begin{description}
\item[\textbf{Random Forest}] is an ensemble of decision trees, where
each tree is trained on a random bootstrap sample of the training data and uses a
random subset of features at each split~\cite{Breiman_2001}. 
Random Forest provides built-in feature
importance scores and is robust to high-dimensional inputs, making it well suited
to the large feature vectors considered in this chapter.
\item[\textbf{XGBoost}] is a gradient boosted tree method that builds an ensemble of trees
sequentially, with each tree trained to correct the errors of the previous ones.
It is known for strong performance on structured data~\cite{xgboost}.
\item[\textbf{CatBoost}] is another gradient boosting tree-based method that has become very popular
due to its strong performance and efficiency~\cite{catboost}.
\item[\textbf{Support Vector Machine}] (SVM) finds the hyperplane that maximizes the margin
between classes in the feature space, using a kernel function to handle non-linear
boundaries~\cite{svm}. In this work, an RBF kernel SVM is considered.
\item[\textbf{Multilayer Perceptron}] (MLP) is a feed-forward neural network with one or
more hidden layers trained by backpropagation, allowing it to model non-linear
relationships between the features and the class labels.
\item[\textbf{$\bm{k}$-Nearest Neighbors}] ($k$-NN) classifies each sample by a majority vote
among its~$k$ closest training samples in feature space; we use~$k = 5$.
\item[\textbf{Logistic Regression}] (LR) is a linear model that estimates class
probabilities directly, and serves as a simple baseline among our classifiers.
\end{description}
These models are evaluated consistently across all three feature extraction tracks 
to allow direct comparison of the feature representations under the same 
classification conditions.

\subsection{Evaluation and Validation Methods}\label{sec:stat_background}

In addition to standard accuracy and F1-score, we employ the following methods
to validate our results and to analyze ensemble behavior.

\begin{description}
\item[\textbf{Bootstrap confidence intervals}] quantify the uncertainty in an accuracy
estimate that is due to the finite size of the test set. The test predictions
are resampled with replacement many times and the accuracy is recomputed on
each resample; the middle~95\%\ of these values forms the confidence interval.
\item[\textbf{McNemar's test}] is a paired significance test for comparing two
classifiers evaluated on the same test samples. It considers only the
discordant pairs, that is, samples that exactly one of the two classifiers
labels correctly, and tests whether one classifier is correct significantly
more often than the other.
\item[\textbf{Diversity measures}] quantify how differently two ensemble members
behave. We report the pairwise disagreement rate, Cohen's~$\kappa$, and the
Yule Q-statistic, where higher disagreement and lower~$\kappa$ or~$Q$
indicate more diverse voters.
\end{description}

\section{Implementation}\label{sec:implementation}

This section describes the complete implementation of our malware classification
pipeline. As discussed above, this work is organized across three feature extraction 
tracks---each producing a different representation of the same malware images---followed 
by various ensembles that combine predictions from one or more of the three tracks.

\begin{figure}[!htb]
\centering
\includegraphics[scale=0.7]{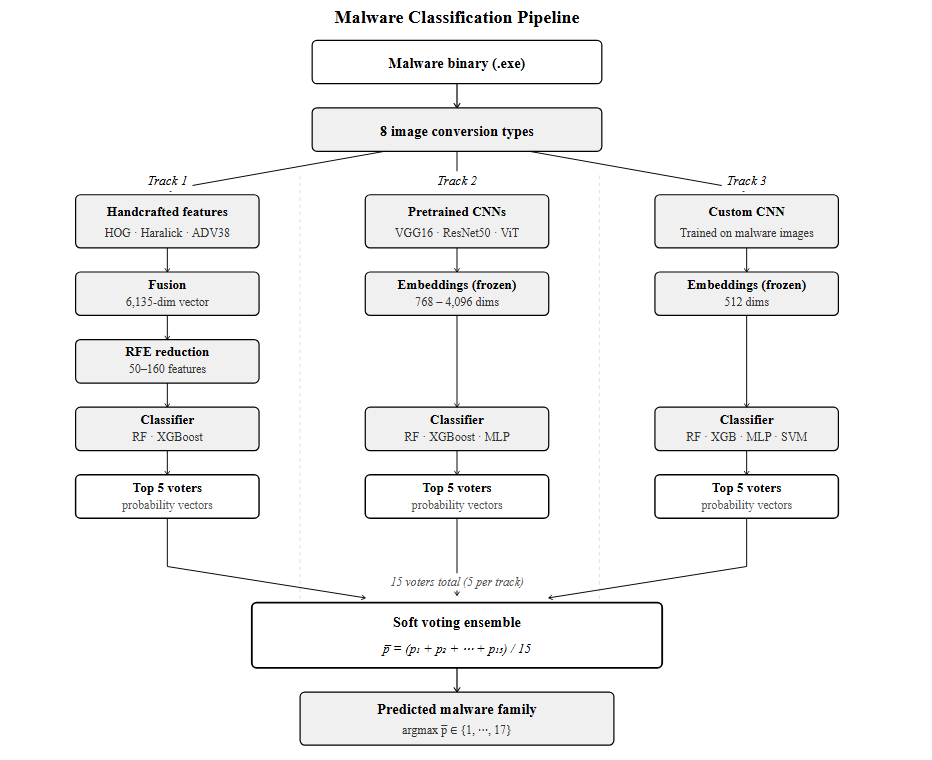}
\caption{Overview of the malware classification pipeline}\label{fig:pipeline}
\end{figure}

The complete classification pipeline is illustrated in
Figure~\ref{fig:pipeline}. The three parallel feature extraction tracks
operate independently on the same malware images, each producing a
set of probability vectors that are combined via soft voting ensembles.

The three feature extraction tracks share the same dataset 
and the same evaluation protocol. 
The held-out test split remains unchanged across
all tracks. Each track is developed and evaluated independently
before being combined in the ensemble. The following sections describe the dataset
and splitting strategy, followed by each feature extraction track, and concluding with 
the ensemble construction technique and evaluation metrics.

\subsection{Dataset and Splitting Strategy}\label{sec:dataset}

The images used in our research originate from the work of Agrawal et al.~\cite{Agrawal_2025}, 
where malware binaries are extracted from the RawMal-TF dataset~\cite{rawmaltf}, and for
each selected sample, eight distinct image representations (as discussed in Section~\ref{sec:conversions})
are generated. The resulting dataset is comprised of~17 malware families, 
with~1,000 malware samples per family, giving us a total of~17,000 samples. 
Since there are eight images per sample, we have a total of~136,000 malware images. 

The~17 malware families in the dataset are
\texttt{Agensla}~\cite{agensla_kaspersky},
\texttt{Androm}~\cite{kasperskyandrom},
\texttt{Convagent}~\cite{convagent},
\texttt{Crypt}~\cite{kasperskycrypt},
\texttt{Crysan}~\cite{kasperskycrysan},
\texttt{DCRat}~\cite{kasperskydcrat},
\texttt{Injuke}~\cite{kasperskyinjuke},
\texttt{Makoob}~\cite{kasperskymokes},
\texttt{Mokes}~\cite{kasperskymokes},
\texttt{Noon}~\cite{kasperskynoon},
\texttt{Remcos}~\cite{kasperskyremcos},
\texttt{Seraph}~\cite{kasperskyseraph},
\texttt{SnakeLogger}~\cite{kasperskysnakelogger},
\texttt{Stealerc}~\cite{kasperskystealerc},
\texttt{Strab}~\cite{kasperskystrab},
\texttt{Taskun}~\cite{kasperskytaskun}, and
\texttt{Zenpak}~\cite{kasperskyzenpak}.
These represent a wide variety of malware types, including \hbox{Trojans}, 
ransomware, encryptors, infostealers, keyloggers, backdoors, RATs, and more.

A single stratified split is applied to the dataset before any feature extraction
is performed. We use a~70-10-10-10 split into training, tuning-validation,
selection-validation, and test sets, so that each family contributes~700 samples to
the training split and~100 samples to each of the three remaining splits. Across
all~17 families this yields~11,900 training samples and~1,700 samples in each of the
tuning-validation, selection-validation, and test splits per conversion type. The
tuning-validation split is used for hyperparameter and feature-count selection; the
selection-validation split is used exclusively to rank voters and choose ensemble
pools; and the test split is held out, untouched, for a single final evaluation. The
same split is applied consistently across all eight conversion types and all three
feature tracks. Because the split is keyed on the SHA-256 hash of each malware binary
(embedded in every filename across all conversions), the identical set of binaries is
guaranteed to appear in the same split regardless of feature type or conversion, and
we verify that every per-voter probability matrix shares the same sample ordering
before averaging. This alignment is critical for the ensemble, where predictions from
models trained on different feature types must correspond to the same test samples.
We use a single stratified split rather than $k$-fold cross-validation, as each fold
would require retraining every configuration across the three tracks and eight
conversion types, including the custom CNNs; split-level uncertainty is instead
quantified with bootstrap confidence intervals and a paired significance test
(Section~\ref{sec:statistics}).

Note that since there are eight images per sample, the number of images per family
is~5,600 in the training split and~800 in each of the tuning-validation,
selection-validation, and test splits. Thus, in total there are~95,200 images in
the training split and~13,600 images in each of the three remaining splits.

Of course, the test split is held out entirely and is never used during model training,
feature selection, or any model selection decision---all design choices are made
based on training and validation performance only. Results are then computed on
the test split.

\subsection{Track~1: Handcrafted Descriptors}\label{sec:track1}

The first track extracts three types of handcrafted descriptors from each malware
image: Histogram of Oriented Gradients (HOG)~\cite{Dalal_2005}, Haralick texture
features~\cite{Haralick_1973}, and the ADV38 statistical features~\cite{Chavda_2020}.
These are extracted independently, concatenated into a single feature vector, and
reduced using a two-stage selection procedure.

\subsubsection{Feature Extraction and Fusion}\label{sec:track1_extraction}

For every image of every conversion type, the three feature sets are extracted and
saved separately for each split. HOG features are extracted using~9 gradient
orientations, $16 \times 16$ pixels per cell, $2 \times 2$ cells per block, and L2-Hys
normalization, producing a~6,084-dimensional vector per image. This dimension follows
from the HOG parameters: a~$224 \times 224$ image with~$16 \times 16$ pixel cells
gives~14 cells per side, yielding~$13 \cdot 13 = 169$ overlapping blocks of~$2\times 2$
cells, each contributing~$2 \cdot 2 \cdot 9 = 36$ values, for a total of
$169 \cdot 36 = 6{,}084$. Haralick features
are computed using the \texttt{mahotas} library on images resized to~$224 \times 224$ 
and converted to grayscale, producing the~13 features originally defined
by Haralick et al.~\cite{Haralick_1973} (contrast, correlation, energy,
homogeneity, entropy, and related measures).
ADV38 features capture metadata, color statistics, texture, shape, and noise
properties, producing~38 features per image.
The three sets are concatenated 
to form a~6,135-dimensional fused vector per image.

\subsubsection{Two-Stage Feature Reduction}\label{sec:track1_reduction}

The 6,135-dimensional vector contains considerable redundancy, particularly
within the HOG block. A two-stage reduction is applied independently for each
conversion type to identify a more compact feature subset.

\begin{description}
\item[\textbf{Stage 1}] (Random Forest Ranking) --- A Random Forest~\cite{Breiman_2001} is
trained on the fused training features. Feature importance scores are used to
rank all~6,135 features, and the top~1,840 features (30\%\ of the
total) are retained. This removes clearly uninformative features
while keeping a broad mix of all three descriptor types.
\item[\textbf{Stage 2}] (Iterative Fraction-Drop Refinement) --- Starting from the top-ranked
subset, an iterative elimination loop is applied:
\begin{enumerate}
    \item Train a Random Forest on the current subset using the training split
    \item Rank features by importance
    \item Drop the lowest-ranked~20\%\ of features
    \item Evaluate macro-averaged F1 on the validation split
    \item Repeat until a minimum of~50 features remain, so that the search
    explores compact subsets while retaining enough capacity to represent
    all three descriptor types
\end{enumerate}
The subset that achieves the highest validation F1 during this process is
saved as the final selected feature set for a given conversion type. Selected features
are applied consistently to train, validation,
and test splits for all downstream experiments.
\end{description}

Final reduced subsets range from 50 features for \grayscale\ to 160
features for \hit. Table~\ref{tab:rfe_subsets} summarizes the composition
of the selected subsets for each image conversion type.

\begin{table}[!htb]
\centering
\caption{RFE-selected feature subsets per conversion type}\label{tab:rfe_subsets}
\begin{adjustbox}{scale=0.85}
\begin{tabular}{lrrcc}
\toprule
\textbf{Conversion} & \textbf{Total} & \textbf{HOG} &
\textbf{Haralick} & \textbf{ADV38} \\
\midrule
\grayscale          &   50 & 26 &  \zz9 & 15 \\
\entropyHilbert            &   67 &  38 & 11 & 18 \\
\byteclass          &   83 &  53 & 13 & 17 \\
\byteclassHilbert\  & 83 &  62 & 11 & 10 \\
\hit                &  160 &  118 & 13 & 29 \\
\spiral             &  388 &  352 & 13 & 23 \\
\bigramCartesian  &  484 &  449 & 13 & 22 \\
\bigramPolar      & 1,178 & 1,141 & 13 & 24 \\
\bottomrule
\end{tabular}
\end{adjustbox}
\end{table}

\subsubsection{Classifiers}\label{sec:track1_classifiers}

Random Forest with~200 estimators and XGBoost with~200 estimators are evaluated
on the selected features for each conversion type. All features are standardized
using a \texttt{StandardScaler} fitted on the training split only.

\subsection{Track~2: Embeddings from Pretrained Models}\label{sec:track2}

For the second track, three neural networks pretrained on ImageNet
are used as fixed feature extractors: ResNet50~\cite{He_2016},
VGG16~\cite{Simonyan_2015}, and ViT-B/16~\cite{Dosovitskiy_2021}. Each network is
loaded with pretrained weights and used in inference mode with its classification
head removed. A malware image is passed through the network, and the output of the
penultimate layer is saved as a dense embedding vector.

\subsubsection{Image Preprocessing}\label{sec:track2_preprocess}

Each malware image is converted to three-channel RGB format, as required by the
pretrained models, and resized to~$224 \times 224$. Images are then normalized
using the per-channel mean and standard deviation
$$
    \boldsymbol{\mu} = (0.485,\, 0.456,\, 0.406) \mbox{\ \ and\ \ }
    \boldsymbol{\sigma} = (0.229,\, 0.224,\, 0.225)
$$
which correspond to the ImageNet training statistics used during pretraining of all
three architectures~\cite{Deng_2009}.

\subsubsection{Model Architectures and Embedding Extraction}\label{sec:track2_arch}

For each of the three pretrained networks, the classification head is
removed to produce a dense embedding rather than a class logit. ResNet50
generates a~2,048-dimensional embedding through its global average pooling
layer. VGG16 generates a~4,096-dimensional embedding through its penultimate
fully connected layer. ViT generates a~768-dimensional embedding from the
class token output by the transformer encoder, since the image is divided
into~$16 \times 16$ pixel patches, producing~196 patch tokens plus one class
token for a~$224 \times 224$ input.
Table~\ref{tab:cnn_embeddings} summarizes the three architectures and their
embedding dimensions.

\begin{table}[!htb]
\centering
\caption{Pretrained CNN architectures used for embedding extraction}\label{tab:cnn_embeddings}
\begin{adjustbox}{scale=0.85}
\begin{tabular}{llr}
\toprule
\multirow{2}{*}{\textbf{Model}} & \multirow{2}{*}{\textbf{Output layer}} & \multicolumn{1}{c}{\textbf{Embedding}} \\
     &  &  \multicolumn{1}{c}{\textbf{dimension}} \\
\midrule
ResNet50   & Global average pooling & 2{,}048\hspace*{0.1in} \\
VGG16      & Penultimate FC layer   & 4{,}096\hspace*{0.1in} \\
ViT   & Class token            & 768\hspace*{0.1in} \\
\bottomrule
\end{tabular}
\end{adjustbox}
\end{table}

\subsubsection{Classifier Training and Evaluation}\label{sec:track2_classifiers}

Feature extraction is performed once per model--conversion combination,
yielding~$3 \times 8 = 24$ embedding sets. Six classifiers are then
trained and evaluated on each set, resulting in~$6 \times 24 = 144$ experiments.
The six classifiers are: Random Forest, XGBoost, SVM (RBF kernel),
MLP, $k$-NN ($k=5$), and Logistic Regression.
All features are standardized using a \texttt{StandardScaler} fitted on the
training split only, and evaluation is performed on the held-out test split.

\subsection{Track~3: Custom CNN Embeddings}\label{sec:track3}

The third track trains a custom CNN from scratch on malware images. The motivation
here is that pretrained models are optimized for natural images and may not capture
the structural patterns specific to malware visualizations.

\subsubsection{Architecture}\label{sec:track3_arch}

\begin{figure}[!htb]
\centering
\includegraphics[scale=0.275]{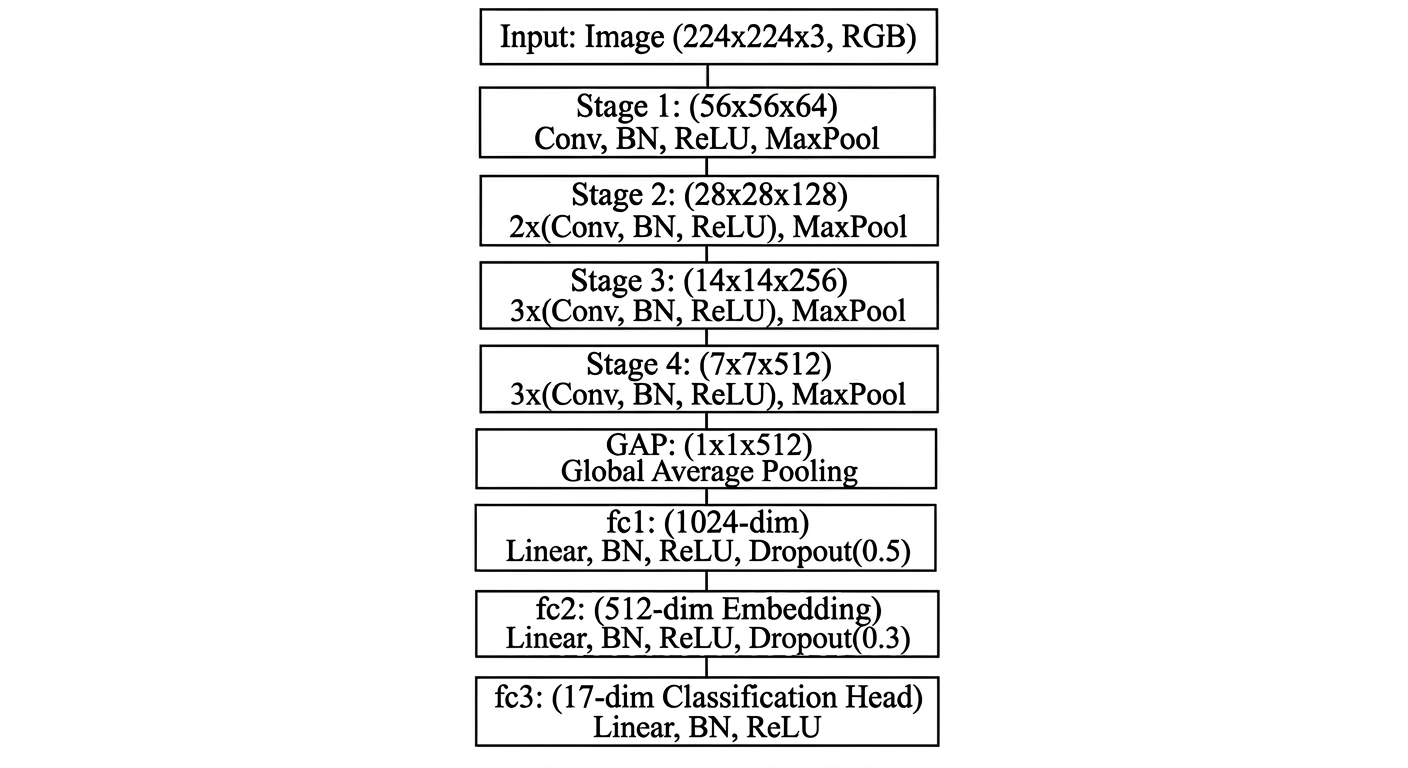}
\caption{Custom CNN architecture}\label{fig:malwarecnn_arch}
\end{figure}

Our custom CNN architecture follows the progressive channel-doubling design
used in established image recognition models~\cite{He_2016,Simonyan_2015}, and
consists of four convolutional stages, global average pooling, and three fully connected layers.
This standard design was chosen because it has proven effective for learning
hierarchical visual features, and because it enables a direct comparison with
the pretrained CNN embeddings under the same experimental conditions---the
512-dimensional embedding produced by the penultimate layer matches the order
of magnitude of the pretrained embeddings.
Figure~\ref{fig:malwarecnn_arch} illustrates the full architecture.

Each convolutional stage applies one or more~$3 \times 3$ convolutions, followed by batch
normalization, ReLU activation, and~$2 \times 2$ max pooling. Channel depth
increases from~64 in Stage~1 to~512 in Stage~4, while spatial dimensions reduce
from~$224 \times 224$ to~$7 \times 7$. A global average pooling layer then
produces a flat~512-dimensional vector, which passes through two fully connected
layers before reaching the classification head. The penultimate layer (\texttt{fc2})
is the source of the feature embedding; the classification head (\texttt{fc3}) is
removed after training. Table~\ref{tab:custom_arch} summarizes the architecture.

\begin{table}[!htb]
\centering
\caption{CNN architecture summary}\label{tab:custom_arch}
\begin{adjustbox}{scale=0.85}
\begin{tabular}{lccl}
\toprule
\textbf{Stage} & \textbf{Output} & \textbf{Dimension} & \textbf{Operations} \\
\midrule
Input   & $224\times224$ & \zz\zz3   & --- \\
Stage 1 & $56\times56$   & \zz64  & Conv $(7\times7)$, $s=2$, BatchNorm, ReLU, MaxPool \\
Stage 2 & $28\times28$   & 128 & 2 Conv $(3\times 3)$, BatchNorm, ReLU, MaxPool \\
Stage 3 & $14\times14$   & 256 & 3 Conv $(3\times 3)$, BatchNorm, ReLU, MaxPool \\
Stage 4 & $7\times7$     & 512 & 3 Conv $(3\times 3)$, BatchNorm, ReLU, MaxPool \\
GAP     & $1\times1$     & 512 & Global average pooling \\
\texttt{fc1}     & 1,024           & --- & Linear, BatchNorm, ReLU, Dropout (0.5) \\
\texttt{fc2}     & 512            & --- & Linear, BatchNorm, ReLU, Dropout (0.3) \\
\texttt{fc3}     & 17             & --- & Linear, removed after training \\
\bottomrule
\end{tabular}
\end{adjustbox}
\end{table}

\subsubsection{Training}\label{sec:track3_train}

The network is trained separately for each of the eight image conversion types.
The training and validation splits are merged into a single set of 15,300
samples to maximize the data available for representation learning.  
Images are preprocessed identically to Track~2.
Using the same normalization as the pretrained models ensures consistent input
scaling across all three tracks. Training uses the
Adam optimizer with cosine annealing and cross-entropy loss for~80 epochs. The
checkpoint with the highest validation accuracy is retained.
Table~\ref{tab:cnn_training} lists the full training configuration.

\begin{table}[!htb]
\centering
\caption{CNN training configuration}\label{tab:cnn_training}
\begin{adjustbox}{scale=0.85}
\begin{tabular}{ll}
\toprule
\textbf{Hyperparameter} & \textbf{Value} \\
\midrule
Optimizer        & Adam \\
Learning rate    & $1 \times 10^{-3}$ \\
Weight decay     & $1 \times 10^{-4}$ \\
LR schedule      & Cosine annealing ($T_{\max} = 80$) \\
Loss             & Cross-entropy \\
Epochs           & 80 \\
Batch size       & 32 \\
Weight init      & He initialization \\
Training samples & 15{,}300 (train/validation merged) \\
\bottomrule
\end{tabular}
\end{adjustbox}
\end{table}

The configuration follows standard practices for training CNNs from scratch.
Adam is used with a learning rate of~$1\times10^{-3}$ and a weight decay
of~$1\times10^{-4}$ as a light regularizer. Cosine annealing gradually
reduces the learning rate without requiring manual step scheduling.
He initialization~\cite{He_2015_init} is applied to maintain consistent
gradient scale with ReLU activations.

\subsubsection{Embedding Extraction and Classifier Training}\label{sec:track3_classifiers}

After training, the final layer (\texttt{fc3}) is removed and the network is used in inference
mode. The ReLU-activated output of the penultimate layer (\texttt{fc2}) is extracted as the
512-dimensional embedding for each image. Embeddings are saved separately for
the training and test splits. For downstream classifier training, only the
13,600 original training samples are used, to maintain consistency with the
other two tracks.

Five classifiers are evaluated on each of the eight feature sets, giving~40
experiments in total. These classifiers are Random Forest, 
XGBoost, SVM (RBF), MLP, and CatBoost, which are described in
Section~\ref{sec:classifiers}. CatBoost is included in this track
in place of LR and KNN, as its ordered boosting is well-suited to the
compact~512-dimensional embedding space.
All models are evaluated on the held-out test split using accuracy and
macro-averaged F1-score.

\subsection{Ensemble Construction}\label{sec:ensemble_construction}

For our ensemble, we combine the probability outputs of multiple classifiers via
soft voting, evaluating a range of pool sizes and compositions rather than fixing
a single voter set in advance. This lets us measure directly whether combining
configurations from multiple feature tracks provides an advantage over the best
single-track pool, rather than assuming it.
Several pool configurations are evaluated to explore the sensitivity of the outcome
to this choice---these include single-track pools, global top-$k$ pools across all
tracks, and balanced per-track pools (Section~\ref{sec:voter_selection}).
An exhaustive search over all
$2^{15}-1 = 32{,}767$ possible voter subsets was not performed. As shown in
Section~\ref{sec:results_ensemble}, accuracy varies only modestly with pool size
and composition. All voters are
trained on the training split only, and the evaluation is performed on the held-out test split.

\subsubsection{Voter Selection}\label{sec:voter_selection}

We rank every track configuration by its accuracy on the selection-validation
split and evaluate a range of voting pools: 
single-track pools of the fifteen best voters in each track, global pools of the
top~$k$ voters across all tracks, and balanced pools of the top~$k$ voters from
each track. The test set is never used for ranking or pool selection. The
best-performing pool, listed in Table~\ref{tab:voters}, consists of the fifteen
top-ranked handcrafted configurations.

\begin{table}[!htb]
\centering
\caption{Voters in the headline pool (top 15 by validation-selection accuracy)}\label{tab:voters}
\begin{adjustbox}{scale=0.85}
\begin{tabular}{cllc}
\toprule
\multirow{2}{*}{\!\!\!\!\!\!\!\textbf{Track}} & \multicolumn{1}{c}{\hspace*{-0.25in}\textbf{Image}} 
	& \multirow{2}{*}{\textbf{Classifier\ \ \ }} & \textbf{Feature} \\
     &\multicolumn{1}{c}{\hspace*{-0.25in}\textbf{conversion}} &  & \textbf{dimension} \\
\midrule
Handcrafted\ \ \         & \hit                         & Random Forest      & 103 \\
Handcrafted\ \ \         & \grayscale                   & Random Forest      & \zz50 \\
Handcrafted\ \ \         & \byteclassHilbert            & XGBoost & 103 \\
Handcrafted\ \ \         & \grayscale                   & XGBoost & \zz50 \\
Handcrafted\ \ \         & \hit                         & XGBoost & 103 \\
Handcrafted\ \ \         & \entropyHilbert              & Random Forest      & \zz54 \\
Handcrafted\ \ \         & \byteclassHilbert            & Random Forest      & 103 \\
Handcrafted\ \ \         & \byteclass                   & Random Forest      & 103 \\
Handcrafted\ \ \         & \byteclass                   & XGBoost & 103 \\
Handcrafted\ \ \         & \entropyHilbert              & XGBoost & \zz54 \\
Handcrafted\ \ \         & \bigramCartesian             & Random Forest      & 249 \\
Handcrafted\ \ \         & \spiral                      & Random Forest      & 128 \\
Handcrafted\ \ \         & \bigramCartesian             & XGBoost & 249 \\
Handcrafted\ \ \         & \bigramPolar                 & Random Forest      & 484 \\
Handcrafted\ \ \         & \spiral                      & XGBoost & 128 \\
\bottomrule
\end{tabular}
\end{adjustbox}
\end{table}

\subsubsection{Soft Voting}\label{sec:soft_voting_impl}

Each voter is trained independently on its own feature representation. Before
training, all features are standardized using a \texttt{StandardScaler} fitted
on the training split only. At inference time, each voter produces 
a~17-dimensional probability distribution over the malware families for each test
sample. The final prediction is the class with the highest average probability
across all~15 voters, which is given by
$$
  \widehat{y} = \displaystyle\argmax_{c}\; \frac{1}{K} \sum_{k=1}^{K} p_{k,c}
$$
where~$K = 15$ is the number of voters and~$p_{k,c}$ is the probability
assigned by voter~$k$ to class~$c$.

To isolate the contribution of each track, multiple voting pool configurations
are evaluated alongside the full~15-voter ensemble. The single-track and
reduced pools serve as ablations. Table~\ref{tab:ensemble_pools} lists all
experimentally evaluated configurations.

\begin{table}[!htb]
\centering
\caption{Voting pool configurations evaluated (all ranked on the validation-selection split)}\label{tab:ensemble_pools}
\begin{adjustbox}{scale=0.85}
\begin{tabular}{lc}
\toprule
\multicolumn{1}{l}{\textbf{Pool}\ \ \ } & \textbf{Voters} \\
\midrule
Handcrafted track, top 15  & 15 \\
Pretrained track, top 15   & 15 \\
Custom CNN track, top 15   & 15 \\
\midrule
Global top 3   &  \zz3 \\
Global top 10  & 10 \\
Global top 15  & 15 \\
Global top 20  & 20 \\
Global top 25  & 25 \\
Global top 30  & 30 \\
\midrule
Top 3 per track   &  \zz9 \\
Top 5 per track   & 15 \\
Top 10 per track  & 30 \\
Top 15 per track  & 45 \\
\bottomrule
\end{tabular}
\end{adjustbox}
\end{table}

\section{Results and Discussion}\label{sec:results}

This section presents the experimental results for each of the three feature
extraction tracks and for our soft voting ensemble. 
Detailed per-configuration results are provided in the appendices: Appendix~A
covers the handcrafted feature analysis, Appendix~B provides per-configuration
results for the pretrained and custom CNN tracks together with the per-class
ensemble report, and Appendix~C provides the global voter ranking and ensemble
pool results.

\subsection{Track 1: Handcrafted Features}\label{sec:results_noncnn}

\begin{figure}[!htb]
\centering
\begin{tikzpicture}[scale=0.60]
    \begin{axis}[
        width=13cm,
        height=13cm,
	colormap={bluewhite}{color=(white) rgb255=(100,149,237)},
        xticklabels={
        Agensla,
        Androm,
        Convagent,
        Crypt,
        Crysan,
        DCRat,
        Injuke,
        Makoob,
        Mokes,
        Noon,
        Remcos,
        Seraph,
        SnakeLogger,
        Stealerc,
        Strab,
        Taskun,
        Zenpak
        },
        xtick={0,...,16},
        xtick style={draw=none},
	xticklabel style={scale=1.25,anchor=east,rotate=60,yshift=-5pt,font=\tt},
        yticklabels={
        Agensla,
        Androm,
        Convagent,
        Crypt,
        Crysan,
        DCRat,
        Injuke,
        Makoob,
        Mokes,
        Noon,
        Remcos,
        Seraph,
        SnakeLogger,
        Stealerc,
        Strab,
        Taskun,
        Zenpak
        },
        ytick={0,...,16},
        ytick style={draw=none},
        enlargelimits=false,
        yticklabel style={scale=1.25,font=\tt},
        colorbar,
        colorbar style={
            ytick={0,20,40,60,80,100},
            yticklabels={0,20,40,60,80,100},
            yticklabel={\pgfmathprintnumber\tick},
            yticklabel style={
            		scale=1.25,
            		/pgf/number format/fixed,
			/pgf/number format/fixed zerofill,
			/pgf/number format/precision=0}
        },
        point meta min=0,
        point meta max=100,
        nodes near coords={\pgfmathprintnumber\pgfplotspointmeta},
        nodes near coords black white/.style={
            small value/.style={
                yshift=-7pt,
                text=black,
                /pgf/number format/fixed,
                /pgf/number format/precision=0,
                /pgf/number format/zerofill=true,
                scale=0.95,
            },
            large value/.style={
                yshift=-7pt,
                text=white,
                /pgf/number format/fixed,
                /pgf/number format/precision=0,
                /pgf/number format/zerofill=true,
                scale=0.95,
            },
            every node near coord/.style={
                check for zero/.code={
                    \pgfmathfloatifflags{\pgfplotspointmeta}{0}{
                        \pgfkeys{/tikz/coordinate}
                    }{
                        \begingroup
                        \pgfkeys{/pgf/fpu}
                        \pgfmathparse{\pgfplotspointmeta<#1}
                        \global\let\result=\pgfmathresult
                        \endgroup
                        %
                        %
                        \pgfmathfloatcreate{1}{1.0}{0}
                        \let\ONE=\pgfmathresult
                        \ifx\result\ONE
                            \pgfkeysalso{/pgfplots/small value}
                        \else
                            \pgfkeysalso{/pgfplots/large value}
                        \fi
                    }
                },
                check for zero,
            },
        },
        nodes near coords black white=50,
    ]
\addplot[
  matrix plot,
  mesh/cols=17,
  point meta=explicit,
  draw=gray
] table [x=x, y=y, meta=C] {
x y C
0 0 59
1 0 3
2 0 0
3 0 4
4 0 1
5 0 0
6 0 2
7 0 0
8 0 0
9 0 8
10 0 0
11 0 7
12 0 1
13 0 0
14 0 0
15 0 15
16 0 0
0 1 1
1 1 65
2 1 1
3 1 2
4 1 1
5 1 0
6 1 3
7 1 7
8 1 0
9 1 3
10 1 2
11 1 6
12 1 0
13 1 5
14 1 0
15 1 3
16 1 1
0 2 0
1 2 0
2 2 70
3 2 0
4 2 0
5 2 0
6 2 9
7 2 1
8 2 1
9 2 0
10 2 1
11 2 1
12 2 0
13 2 5
14 2 8
15 2 0
16 2 4
0 3 6
1 3 0
2 3 2
3 3 68
4 3 0
5 3 0
6 3 3
7 3 0
8 3 0
9 3 1
10 3 1
11 3 12
12 3 0
13 3 1
14 3 0
15 3 5
16 3 1
0 4 1
1 4 0
2 4 1
3 4 1
4 4 83
5 4 0
6 4 2
7 4 0
8 4 0
9 4 0
10 4 0
11 4 8
12 4 0
13 4 3
14 4 0
15 4 1
16 4 0
0 5 0
1 5 0
2 5 0
3 5 0
4 5 0
5 5 99
6 5 0
7 5 0
8 5 0
9 5 0
10 5 0
11 5 1
12 5 0
13 5 0
14 5 0
15 5 0
16 5 0
0 6 4
1 6 1
2 6 11
3 6 3
4 6 1
5 6 0
6 6 60
7 6 0
8 6 0
9 6 1
10 6 2
11 6 9
12 6 0
13 6 3
14 6 2
15 6 0
16 6 3
0 7 0
1 7 0
2 7 0
3 7 0
4 7 0
5 7 0
6 7 0
7 7 99
8 7 0
9 7 0
10 7 0
11 7 0
12 7 0
13 7 0
14 7 1
15 7 0
16 7 0
0 8 0
1 8 1
2 8 1
3 8 0
4 8 0
5 8 0
6 8 0
7 8 0
8 8 81
9 8 0
10 8 0
11 8 1
12 8 0
13 8 8
14 8 0
15 8 0
16 8 8
0 9 7
1 9 1
2 9 0
3 9 3
4 9 1
5 9 0
6 9 3
7 9 1
8 9 0
9 9 67
10 9 4
11 9 2
12 9 0
13 9 2
14 9 1
15 9 8
16 9 0
0 10 1
1 10 0
2 10 0
3 10 1
4 10 1
5 10 0
6 10 1
7 10 0
8 10 0
9 10 2
10 10 80
11 10 6
12 10 1
13 10 0
14 10 1
15 10 5
16 10 1
0 11 0
1 11 0
2 11 0
3 11 0
4 11 1
5 11 0
6 11 1
7 11 0
8 11 0
9 11 2
10 11 0
11 11 94
12 11 1
13 11 0
14 11 0
15 11 1
16 11 0
0 12 2
1 12 0
2 12 0
3 12 1
4 12 1
5 12 1
6 12 0
7 12 0
8 12 0
9 12 1
10 12 0
11 12 7
12 12 83
13 12 0
14 12 2
15 12 2
16 12 0
0 13 0
1 13 0
2 13 3
3 13 3
4 13 0
5 13 0
6 13 3
7 13 0
8 13 8
9 13 0
10 13 1
11 13 3
12 13 0
13 13 69
14 13 7
15 13 1
16 13 2
0 14 0
1 14 0
2 14 0
3 14 1
4 14 0
5 14 0
6 14 0
7 14 0
8 14 1
9 14 0
10 14 3
11 14 1
12 14 0
13 14 5
14 14 89
15 14 0
16 14 0
0 15 7
1 15 0
2 15 0
3 15 1
4 15 0
5 15 0
6 15 1
7 15 0
8 15 0
9 15 5
10 15 3
11 15 1
12 15 0
13 15 0
14 15 0
15 15 82
16 15 0
0 16 0
1 16 0
2 16 1
3 16 0
4 16 0
5 16 0
6 16 0
7 16 0
8 16 16
9 16 0
10 16 0
11 16 0
12 16 0
13 16 6
14 16 2
15 16 0
16 16 75
};
\end{axis}
\end{tikzpicture}
\caption{Confusion matrix for the best handcrafted configuration 
(\hit\ + Random Forest)}\label{fig:handcrafted_cm}
\end{figure}
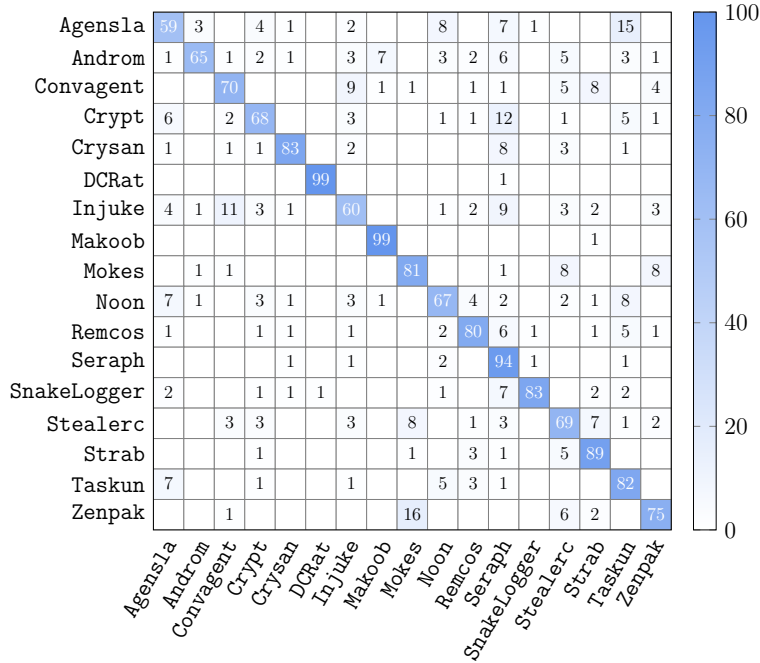

The handcrafted feature track evaluates Random Forest and XGBoost classifiers
trained on the RFE-selected fused feature subsets across all eight conversion
types. Table~\ref{tab:handcrafted_best} summarizes the best result per conversion
type, while Figure~\ref{fig:handcrafted_cm} shows the confusion matrix for this configuration on the test set.
The best result for this track is~77.8\% accuracy, achieved by the
\hit\  image type with Random Forest trained on~103 selected
features. This represents a reduction of more than~98\% from the 
original feature vector of size~6,135, while still outperforming the HOG-only baseline
reported in Agrawal et al.~\cite{Agrawal_2025}.

\begin{table}[!htb]
\centering
\caption{Best handcrafted feature result per conversion type}\label{tab:handcrafted_best}
\begin{adjustbox}{scale=0.85}
\begin{tabular}{lccccc}
\toprule
\textbf{Conversion} & \textbf{Classifier} & \textbf{Features} & \textbf{Accuracy} & \textbf{Precision} & \textbf{F1} \\
\midrule
\hit  & Random Forest  & 103 & 0.7782 & 0.7893 & 0.7781 \\
\grayscale  & Random Forest  & \zz50 & 0.7759 & 0.7855 & 0.7762 \\
\byteclassHilbert  & Random Forest  & 103 & 0.7565 & 0.7700 & 0.7575 \\
\entropyHilbert  & Random Forest  & \zz54 & 0.7541 & 0.7628 & 0.7539 \\
\byteclass  & Random Forest  & 103 & 0.7518 & 0.7625 & 0.7522 \\
\spiral  & Random Forest  & 128 & 0.6912 & 0.6974 & 0.6880 \\
\bigramPolar  & XGBoost  & 484 & 0.6788 & 0.6866 & 0.6781 \\
\bigramCartesian  & Random Forest  & 249 & 0.6765 & 0.6975 & 0.6765 \\
\bottomrule
\end{tabular}
\end{adjustbox}
\end{table}

The RFE-selected subsets show a consistent pattern across all conversion types.
That is, HOG features account for approximately~60\%\ to over~97\%\ of retained features,
reflecting the importance of local gradient structure for distinguishing malware
families. Haralick features are present in every selected subset. ADV38 features,
while the smallest contributor by count, are retained across all conversion types,
confirming that they provide complementary information not captured by HOG or Haralick alone. 

\subsection{Track 2: Pretrained Neural Network Embeddings}\label{sec:results_pretrained}

A total of~144 experiments were conducted across three pretrained neural network models, 
eight conversion types, and six classifiers. Table~\ref{tab:pretrained_best} lists
the top ten results by test accuracy. 

\begin{table}[!htb]
\centering
\caption{Top ten pretrained CNN results on the test set}\label{tab:pretrained_best}
\begin{adjustbox}{scale=0.85}
\begin{tabular}{llccc}
\toprule
\textbf{Model / Conversion} & \textbf{Classifier} & \textbf{Accuracy} & \textbf{Precision} & \textbf{F1} \\
\midrule
ResNet50 / \grayscale & Random Forest & 0.7382 & 0.7472 & 0.7378 \\
ViT / \byteclass & Random Forest & 0.7347 & 0.7455 & 0.7353 \\
VGG16 / \grayscale & Random Forest & 0.7347 & 0.7422 & 0.7336 \\
VGG16 / \byteclass & Random Forest & 0.7347 & 0.7468 & 0.7353 \\
VGG16 / \byteclass & XGBoost & 0.7306 & 0.7401 & 0.7310 \\
ViT / \byteclass & XGBoost & 0.7259 & 0.7332 & 0.7257 \\
ViT / \grayscale & Random Forest & 0.7253 & 0.7353 & 0.7247 \\
ResNet50 / \byteclass & Random Forest & 0.7235 & 0.7332 & 0.7237 \\
ResNet50 / \hit & Random Forest & 0.7235 & 0.7396 & 0.7246 \\
ResNet50 / \grayscale & XGBoost & 0.7218 & 0.7307 & 0.7226 \\
\bottomrule
\end{tabular}
\end{adjustbox}
\end{table}

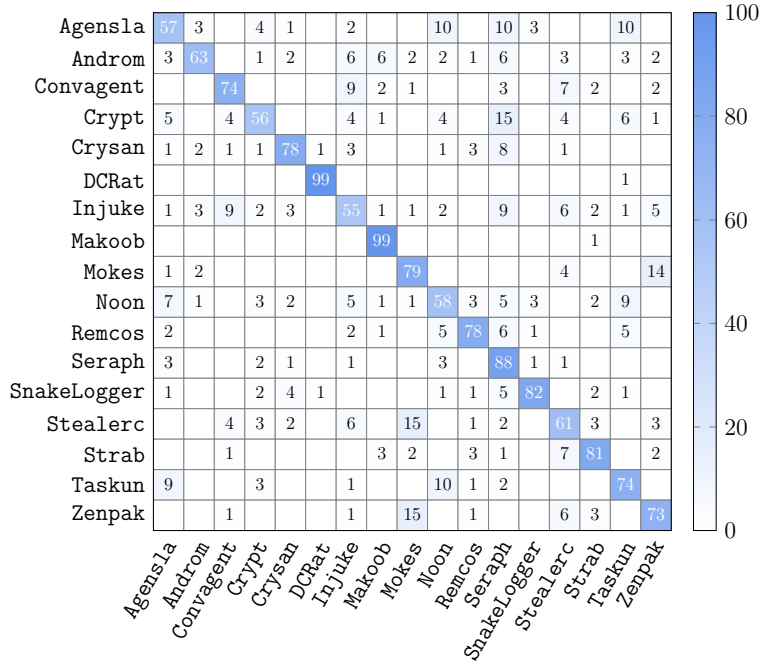
\begin{figure}[!htb]
\centering
\begin{tikzpicture}[scale=0.60]
    \begin{axis}[
        width=13cm,
        height=13cm,
	colormap={bluewhite}{color=(white) rgb255=(100,149,237)},
        xticklabels={
        Agensla,
        Androm,
        Convagent,
        Crypt,
        Crysan,
        DCRat,
        Injuke,
        Makoob,
        Mokes,
        Noon,
        Remcos,
        Seraph,
        SnakeLogger,
        Stealerc,
        Strab,
        Taskun,
        Zenpak
        },
        xtick={0,...,16},
        xtick style={draw=none},
	xticklabel style={scale=1.25,anchor=east,rotate=60,yshift=-5pt,font=\tt},
        yticklabels={
        Agensla,
        Androm,
        Convagent,
        Crypt,
        Crysan,
        DCRat,
        Injuke,
        Makoob,
        Mokes,
        Noon,
        Remcos,
        Seraph,
        SnakeLogger,
        Stealerc,
        Strab,
        Taskun,
        Zenpak
        },
        ytick={0,...,16},
        ytick style={draw=none},
        enlargelimits=false,
        yticklabel style={scale=1.25,font=\tt},
        colorbar,
        colorbar style={
            ytick={0,20,40,60,80,100},
            yticklabels={0,20,40,60,80,100},
            yticklabel={\pgfmathprintnumber\tick},
            yticklabel style={
            		scale=1.25,
            		/pgf/number format/fixed,
			/pgf/number format/fixed zerofill,
			/pgf/number format/precision=0}
        },
        point meta min=0,
        point meta max=100,
        nodes near coords={\pgfmathprintnumber\pgfplotspointmeta},
        nodes near coords black white/.style={
            small value/.style={
                yshift=-7pt,
                text=black,
                /pgf/number format/fixed,
                /pgf/number format/precision=0,
                /pgf/number format/zerofill=true,
                scale=0.95,
            },
            large value/.style={
                yshift=-7pt,
                text=white,
                /pgf/number format/fixed,
                /pgf/number format/precision=0,
                /pgf/number format/zerofill=true,
                scale=0.95,
            },
            every node near coord/.style={
                check for zero/.code={
                    \pgfmathfloatifflags{\pgfplotspointmeta}{0}{
                        \pgfkeys{/tikz/coordinate}
                    }{
                        \begingroup
                        \pgfkeys{/pgf/fpu}
                        \pgfmathparse{\pgfplotspointmeta<#1}
                        \global\let\result=\pgfmathresult
                        \endgroup
                        %
                        %
                        \pgfmathfloatcreate{1}{1.0}{0}
                        \let\ONE=\pgfmathresult
                        \ifx\result\ONE
                            \pgfkeysalso{/pgfplots/small value}
                        \else
                            \pgfkeysalso{/pgfplots/large value}
                        \fi
                    }
                },
                check for zero,
            },
        },
        nodes near coords black white=50,
    ]
\addplot[
  matrix plot,
  mesh/cols=17,
  point meta=explicit,
  draw=gray
] table [x=x, y=y, meta=C] {
x y C
0 0 57
1 0 3
2 0 0
3 0 4
4 0 1
5 0 0
6 0 2
7 0 0
8 0 0
9 0 10
10 0 0
11 0 10
12 0 3
13 0 0
14 0 0
15 0 10
16 0 0
0 1 3
1 1 63
2 1 0
3 1 1
4 1 2
5 1 0
6 1 6
7 1 6
8 1 2
9 1 2
10 1 1
11 1 6
12 1 0
13 1 3
14 1 0
15 1 3
16 1 2
0 2 0
1 2 0
2 2 74
3 2 0
4 2 0
5 2 0
6 2 9
7 2 2
8 2 1
9 2 0
10 2 0
11 2 3
12 2 0
13 2 7
14 2 2
15 2 0
16 2 2
0 3 5
1 3 0
2 3 4
3 3 56
4 3 0
5 3 0
6 3 4
7 3 1
8 3 0
9 3 4
10 3 0
11 3 15
12 3 0
13 3 4
14 3 0
15 3 6
16 3 1
0 4 1
1 4 2
2 4 1
3 4 1
4 4 78
5 4 1
6 4 3
7 4 0
8 4 0
9 4 1
10 4 3
11 4 8
12 4 0
13 4 1
14 4 0
15 4 0
16 4 0
0 5 0
1 5 0
2 5 0
3 5 0
4 5 0
5 5 99
6 5 0
7 5 0
8 5 0
9 5 0
10 5 0
11 5 0
12 5 0
13 5 0
14 5 0
15 5 1
16 5 0
0 6 1
1 6 3
2 6 9
3 6 2
4 6 3
5 6 0
6 6 55
7 6 1
8 6 1
9 6 2
10 6 0
11 6 9
12 6 0
13 6 6
14 6 2
15 6 1
16 6 5
0 7 0
1 7 0
2 7 0
3 7 0
4 7 0
5 7 0
6 7 0
7 7 99
8 7 0
9 7 0
10 7 0
11 7 0
12 7 0
13 7 0
14 7 1
15 7 0
16 7 0
0 8 1
1 8 2
2 8 0
3 8 0
4 8 0
5 8 0
6 8 0
7 8 0
8 8 79
9 8 0
10 8 0
11 8 0
12 8 0
13 8 4
14 8 0
15 8 0
16 8 14
0 9 7
1 9 1
2 9 0
3 9 3
4 9 2
5 9 0
6 9 5
7 9 1
8 9 1
9 9 58
10 9 3
11 9 5
12 9 3
13 9 0
14 9 2
15 9 9
16 9 0
0 10 2
1 10 0
2 10 0
3 10 0
4 10 0
5 10 0
6 10 2
7 10 1
8 10 0
9 10 5
10 10 78
11 10 6
12 10 1
13 10 0
14 10 0
15 10 5
16 10 0
0 11 3
1 11 0
2 11 0
3 11 2
4 11 1
5 11 0
6 11 1
7 11 0
8 11 0
9 11 3
10 11 0
11 11 88
12 11 1
13 11 1
14 11 0
15 11 0
16 11 0
0 12 1
1 12 0
2 12 0
3 12 2
4 12 4
5 12 1
6 12 0
7 12 0
8 12 0
9 12 1
10 12 1
11 12 5
12 12 82
13 12 0
14 12 2
15 12 1
16 12 0
0 13 0
1 13 0
2 13 4
3 13 3
4 13 2
5 13 0
6 13 6
7 13 0
8 13 15
9 13 0
10 13 1
11 13 2
12 13 0
13 13 61
14 13 3
15 13 0
16 13 3
0 14 0
1 14 0
2 14 1
3 14 0
4 14 0
5 14 0
6 14 0
7 14 3
8 14 2
9 14 0
10 14 3
11 14 1
12 14 0
13 14 7
14 14 81
15 14 0
16 14 2
0 15 9
1 15 0
2 15 0
3 15 3
4 15 0
5 15 0
6 15 1
7 15 0
8 15 0
9 15 10
10 15 1
11 15 2
12 15 0
13 15 0
14 15 0
15 15 74
16 15 0
0 16 0
1 16 0
2 16 1
3 16 0
4 16 0
5 16 0
6 16 1
7 16 0
8 16 15
9 16 0
10 16 1
11 16 0
12 16 0
13 16 6
14 16 3
15 16 0
16 16 73
};
\end{axis}
\end{tikzpicture}
\caption{Confusion matrix for the best pretrained neural network configuration 
(ResNet50 / \grayscale\ + Random Forest)}\label{fig:pretrained_cm}
\end{figure}

The best result in this track is~73.8\%\ accuracy, achieved by ResNet50 embeddings
from the \grayscale\ conversion with a Random Forest classifier. VGG16 follows
closely at~73.5\%\ for the same image conversion, while ViT reaches~73.5\%\ on the
\byteclass\ conversion. All of the top results use
Random Forest as the classifier. Figure~\ref{fig:pretrained_cm} shows the confusion
matrix for the best configuration (ResNet50 with \grayscale\ + Random Forest)
on the test set.

\subsubsection{Effect of Classifier}\label{sec:pretrained_classifier}

Random Forest is the strongest classifier across all 
pretrained neural network models and conversion
types, with an average accuracy of~67.6\%. XGBoost is competitive with
an average accuracy of~66.7\%.
MLP performs only slightly worse, at~65.6\%, while SVM is a clear outlier, averaging 
only~49.7\%\ accuracy across all configurations. The poor SVM performance is consistent across
all three pretrained models and likely results from the high-dimensional embedding spaces
(up to~4,096 for VGG16), which makes kernel methods computationally expensive and
less effective without careful tuning. Logistic Regression and KNN both underperform
Random Forest and XGBoost across all settings. CatBoost is not included in this
track, as it is used only in the custom CNN track
(Section~\ref{sec:results_custom}), where the lower-dimensional embeddings
make it a more practical choice.

\subsubsection{Effect of Pretrained Model}\label{sec:pretrained_cnn}

\begin{figure}[!htb]
\centering
\includegraphics[scale=0.425]{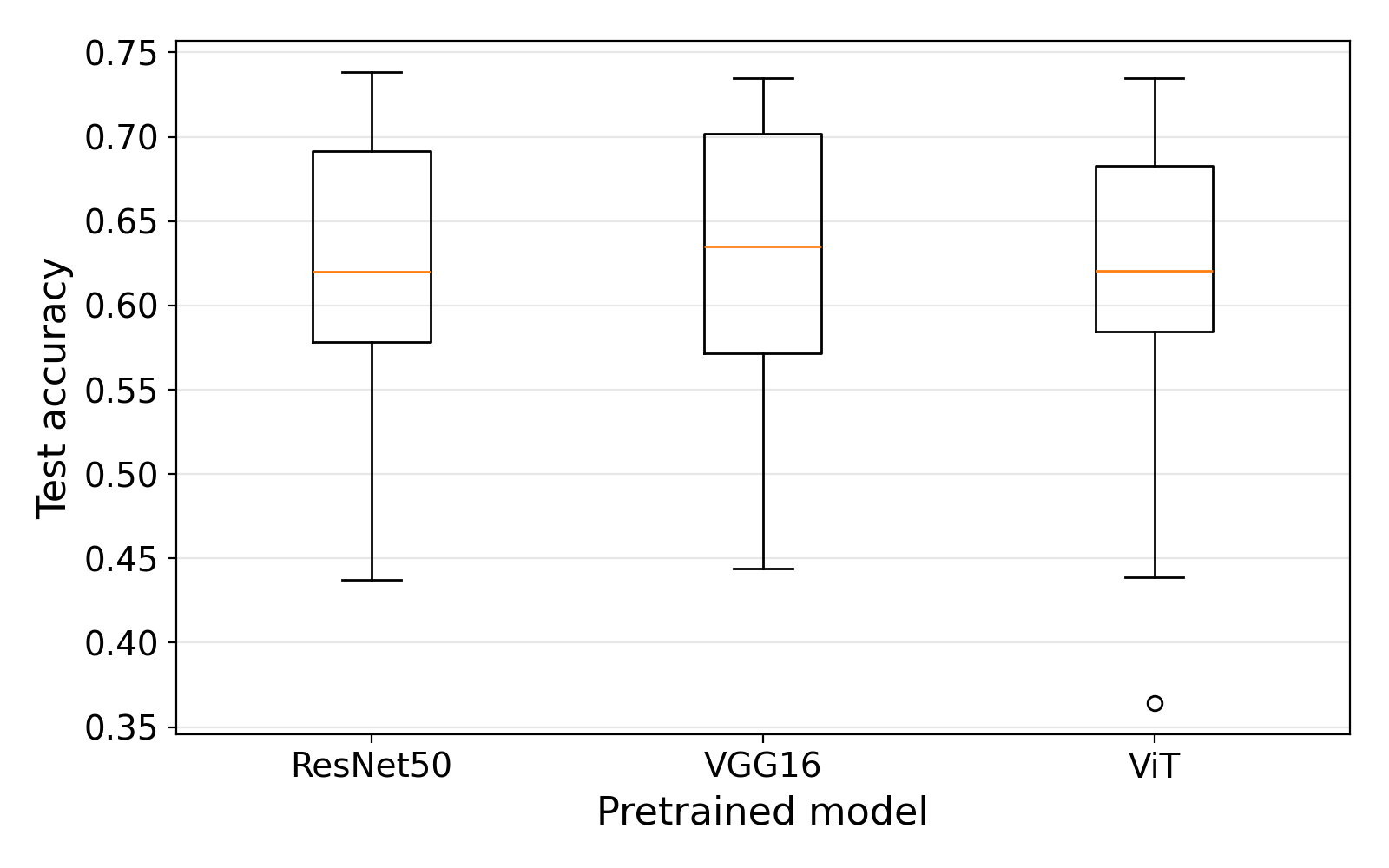}
\caption{Test accuracy distributions (pretrained neural networks)}\label{fig:pretrained_box}
\end{figure}

VGG16 achieves the highest average accuracy at~63.6\%, 
followed by ResNet50 at~62.3\%\ and ViT at~61.7\%. The
performance gap between the three models is small, suggesting that the
choice of classifier and conversion type has a stronger influence
on results than the pretrained architecture itself. All three models show wide
variance across conversion types, as seen in Figure~\ref{fig:pretrained_box},
driven primarily by the weak performance of SVM and the bigram-based
conversions.

\subsubsection{Effect of Conversion Type}\label{sec:pretrained_conversion}

The \grayscale\ and \byteclass\ image conversions consistently produce
the strongest average accuracy at~67.0\%\ and~66.4\%, respectively. 
The \hit\ and \entropyHilbert\ conversions follow at~66.0\%\ 
and~63.9\%, respectively. The \byteclassHilbert\ conversion, which produced the best
result in the handcrafted track, ranks fifth at~63.6\%, suggesting that
the spatial locality preserved by the Hilbert curve benefits handcrafted
gradient features more than dense neural network embeddings. 
The \bigramPolar\ and \spiral\ conversions produce the weakest results at~56.7\%\ 
and~56.3\%\ average accuracy, consistent with the handcrafted track.

\begin{figure}[!htb]
\centering
\includegraphics[scale=0.425]{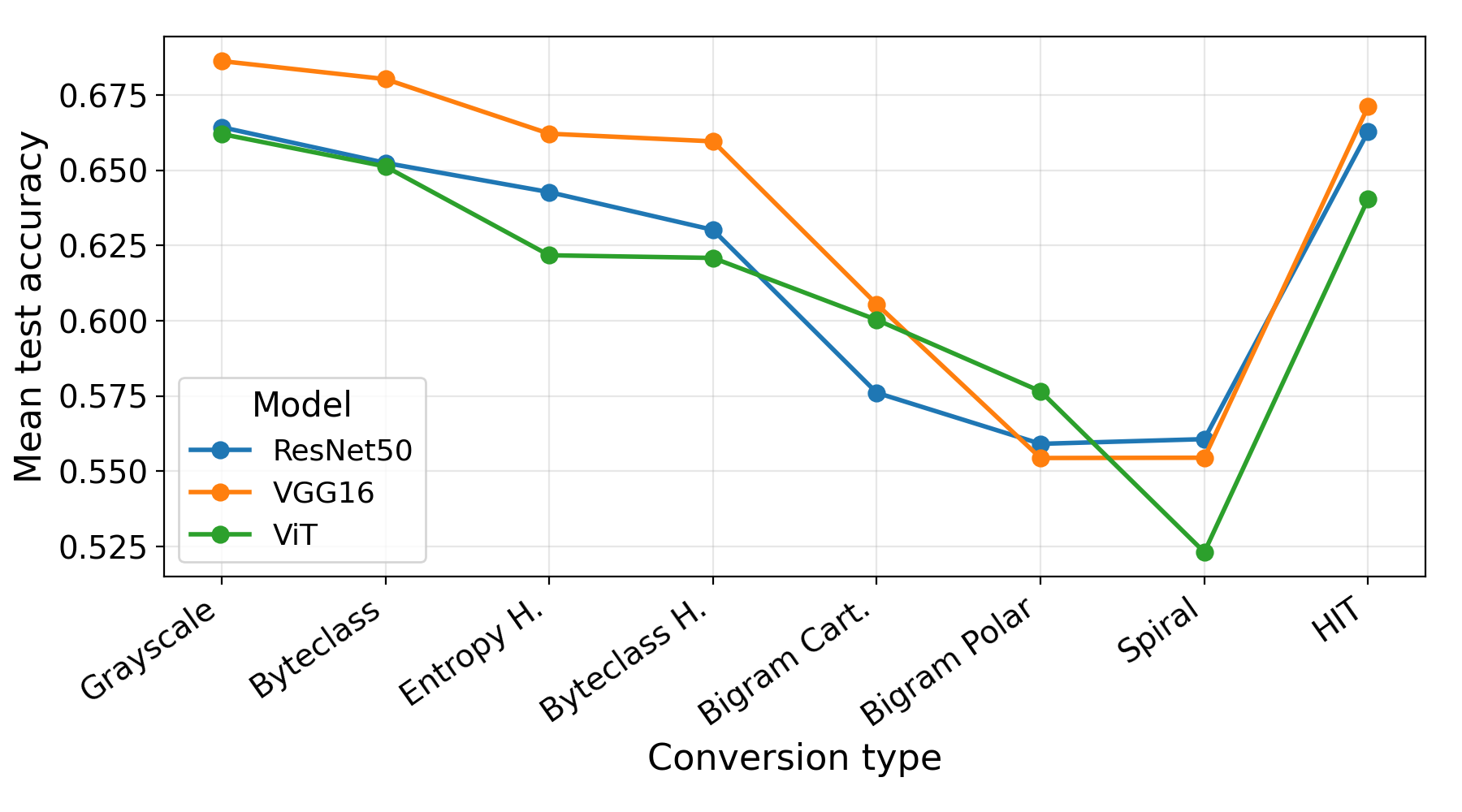}
\caption{Average test accuracy per conversion type (pretrained CNNs)}
\label{fig:pretrained_line}
\end{figure}

\subsubsection{Comparison with Handcrafted Track}\label{sec:pretrained_vs_noncnn}

The best pretrained neural network accuracy of~73.8\%\ is approximately four percentage
points below the best handcrafted result of~77.8\%. This gap is notable
given that VGG16 produces a~4,096-dimensional embedding while the best
handcrafted model uses only~103 selected features.

\subsection{Track 3: Custom CNN Embeddings}\label{sec:results_custom}

A total of~40 experiments were conducted across eight conversion types and
five classifiers. Table~\ref{tab:custom_best} lists the best result per
conversion type. 

\begin{table}[!htb]
\centering
\caption{Best custom CNN result per conversion type}\label{tab:custom_best}
\begin{adjustbox}{scale=0.85}
\begin{tabular}{lccccc}
\toprule
\textbf{Conversion} & \textbf{Classifier} & \textbf{Features} & \textbf{Accuracy} & \textbf{Precision} & \textbf{F1} \\
\midrule
\grayscale & Random Forest & \zz512 & 0.7482 & 0.7491 & 0.7480 \\
\hit & Random Forest & \zz512 & 0.7465 & 0.7501 & 0.7472 \\
\entropyHilbert & Random Forest & \zz512 & 0.7418 & 0.7484 & 0.7433 \\
\byteclassHilbert & Random Forest & \zz512 & 0.7271 & 0.7394 & 0.7290 \\
\bigramCartesian & SVM & \zz512 & 0.7141 & 0.7150 & 0.7133 \\
\bigramPolar & SVM & \zz512 & 0.7094 & 0.7169 & 0.7095 \\
\byteclass & XGBoost & \zz512 & 0.6988 & 0.7105 & 0.6994 \\
\spiral & XGBoost & \zz512 & 0.6941 & 0.6976 & 0.6936 \\
\bottomrule
\end{tabular}
\end{adjustbox}
\end{table}

\begin{figure}[!htb]
\centering
\begin{tikzpicture}[scale=0.60]
    \begin{axis}[
        width=13cm,
        height=13cm,
	colormap={bluewhite}{color=(white) rgb255=(100,149,237)},
        xticklabels={
        Agensla,
        Androm,
        Convagent,
        Crypt,
        Crysan,
        DCRat,
        Injuke,
        Makoob,
        Mokes,
        Noon,
        Remcos,
        Seraph,
        SnakeLogger,
        Stealerc,
        Strab,
        Taskun,
        Zenpak
        },
        xtick={0,...,16},
        xtick style={draw=none},
	xticklabel style={scale=1.25,anchor=east,rotate=60,yshift=-5pt,font=\tt},
        yticklabels={
        Agensla,
        Androm,
        Convagent,
        Crypt,
        Crysan,
        DCRat,
        Injuke,
        Makoob,
        Mokes,
        Noon,
        Remcos,
        Seraph,
        SnakeLogger,
        Stealerc,
        Strab,
        Taskun,
        Zenpak
        },
        ytick={0,...,16},
        ytick style={draw=none},
        enlargelimits=false,
        yticklabel style={scale=1.25,font=\tt},
        colorbar,
        colorbar style={
            ytick={0,20,40,60,80,100},
            yticklabels={0,20,40,60,80,100},
            yticklabel={\pgfmathprintnumber\tick},
            yticklabel style={
            		scale=1.25,
            		/pgf/number format/fixed,
			/pgf/number format/fixed zerofill,
			/pgf/number format/precision=0}
        },
        point meta min=0,
        point meta max=100,
        nodes near coords={\pgfmathprintnumber\pgfplotspointmeta},
        nodes near coords black white/.style={
            small value/.style={
                yshift=-7pt,
                text=black,
                /pgf/number format/fixed,
                /pgf/number format/precision=0,
                /pgf/number format/zerofill=true,
                scale=0.95,
            },
            large value/.style={
                yshift=-7pt,
                text=white,
                /pgf/number format/fixed,
                /pgf/number format/precision=0,
                /pgf/number format/zerofill=true,
                scale=0.95,
            },
            every node near coord/.style={
                check for zero/.code={
                    \pgfmathfloatifflags{\pgfplotspointmeta}{0}{
                        \pgfkeys{/tikz/coordinate}
                    }{
                        \begingroup
                        \pgfkeys{/pgf/fpu}
                        \pgfmathparse{\pgfplotspointmeta<#1}
                        \global\let\result=\pgfmathresult
                        \endgroup
                        %
                        %
                        \pgfmathfloatcreate{1}{1.0}{0}
                        \let\ONE=\pgfmathresult
                        \ifx\result\ONE
                            \pgfkeysalso{/pgfplots/small value}
                        \else
                            \pgfkeysalso{/pgfplots/large value}
                        \fi
                    }
                },
                check for zero,
            },
        },
        nodes near coords black white=50,
    ]
\addplot[
  matrix plot,
  mesh/cols=17,
  point meta=explicit,
  draw=gray
] table [x=x, y=y, meta=C] {
x y C
0 0 55
1 0 7
2 0 0
3 0 4
4 0 2
5 0 0
6 0 5
7 0 0
8 0 0
9 0 9
10 0 6
11 0 1
12 0 2
13 0 1
14 0 0
15 0 8
16 0 0
0 1 3
1 1 71
2 1 0
3 1 1
4 1 2
5 1 0
6 1 5
7 1 4
8 1 1
9 1 1
10 1 1
11 1 3
12 1 2
13 1 4
14 1 0
15 1 0
16 1 2
0 2 0
1 2 2
2 2 73
3 2 0
4 2 1
5 2 0
6 2 8
7 2 1
8 2 1
9 2 0
10 2 0
11 2 1
12 2 0
13 2 4
14 2 4
15 2 0
16 2 5
0 3 7
1 3 1
2 3 3
3 3 62
4 3 3
5 3 0
6 3 7
7 3 0
8 3 1
9 3 3
10 3 1
11 3 9
12 3 0
13 3 2
14 3 0
15 3 1
16 3 0
0 4 0
1 4 3
2 4 1
3 4 4
4 4 80
5 4 0
6 4 0
7 4 0
8 4 0
9 4 3
10 4 1
11 4 5
12 4 1
13 4 2
14 4 0
15 4 0
16 4 0
0 5 1
1 5 0
2 5 0
3 5 0
4 5 0
5 5 99
6 5 0
7 5 0
8 5 0
9 5 0
10 5 0
11 5 0
12 5 0
13 5 0
14 5 0
15 5 0
16 5 0
0 6 4
1 6 0
2 6 9
3 6 6
4 6 1
5 6 0
6 6 63
7 6 0
8 6 1
9 6 1
10 6 2
11 6 4
12 6 1
13 6 5
14 6 0
15 6 2
16 6 1
0 7 0
1 7 2
2 7 0
3 7 0
4 7 0
5 7 0
6 7 0
7 7 97
8 7 0
9 7 0
10 7 0
11 7 0
12 7 0
13 7 0
14 7 1
15 7 0
16 7 0
0 8 0
1 8 2
2 8 1
3 8 0
4 8 0
5 8 0
6 8 1
7 8 0
8 8 74
9 8 0
10 8 0
11 8 0
12 8 0
13 8 3
14 8 0
15 8 0
16 8 19
0 9 10
1 9 2
2 9 0
3 9 3
4 9 2
5 9 1
6 9 3
7 9 1
8 9 0
9 9 64
10 9 5
11 9 2
12 9 0
13 9 0
14 9 1
15 9 6
16 9 0
0 10 1
1 10 2
2 10 0
3 10 2
4 10 0
5 10 0
6 10 0
7 10 0
8 10 0
9 10 4
10 10 80
11 10 6
12 10 2
13 10 1
14 10 1
15 10 1
16 10 0
0 11 4
1 11 1
2 11 0
3 11 2
4 11 5
5 11 0
6 11 5
7 11 0
8 11 0
9 11 6
10 11 4
11 11 71
12 11 1
13 11 1
14 11 0
15 11 0
16 11 0
0 12 1
1 12 0
2 12 0
3 12 0
4 12 0
5 12 0
6 12 0
7 12 0
8 12 0
9 12 1
10 12 4
11 12 2
12 12 88
13 12 0
14 12 3
15 12 1
16 12 0
0 13 1
1 13 3
2 13 4
3 13 1
4 13 2
5 13 0
6 13 9
7 13 0
8 13 8
9 13 0
10 13 0
11 13 1
12 13 0
13 13 63
14 13 3
15 13 0
16 13 5
0 14 2
1 14 0
2 14 2
3 14 0
4 14 0
5 14 0
6 14 0
7 14 1
8 14 0
9 14 0
10 14 2
11 14 0
12 14 0
13 14 10
14 14 78
15 14 0
16 14 5
0 15 6
1 15 1
2 15 0
3 15 2
4 15 0
5 15 0
6 15 0
7 15 0
8 15 0
9 15 6
10 15 2
11 15 1
12 15 5
13 15 0
14 15 0
15 15 77
16 15 0
0 16 0
1 16 0
2 16 1
3 16 0
4 16 0
5 16 0
6 16 2
7 16 0
8 16 14
9 16 0
10 16 0
11 16 0
12 16 0
13 16 4
14 16 2
15 16 0
16 16 77
};
\end{axis}
\end{tikzpicture}
\caption{Confusion matrix for the best custom CNN configuration}\label{fig:custom_cm}
\end{figure}
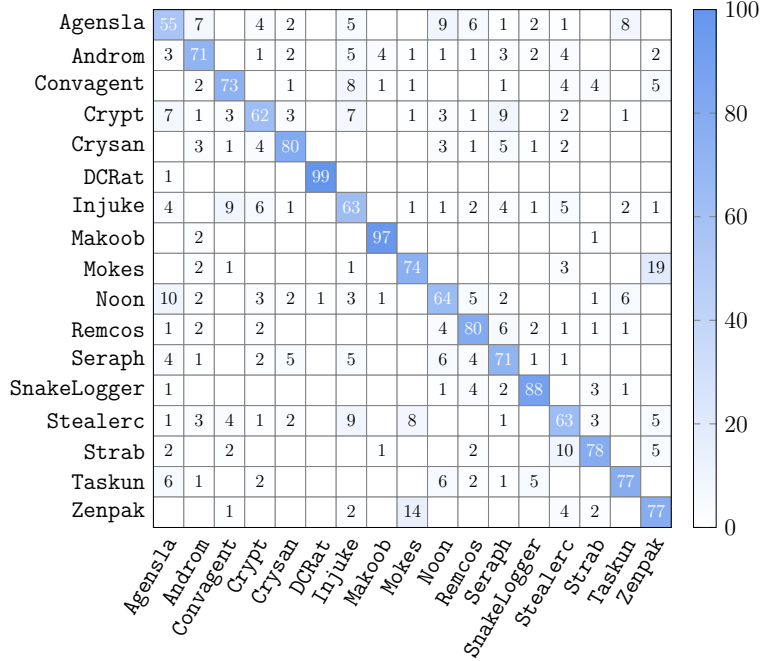

The best result in this track achieves~74.8\% accuracy with the
\grayscale\ conversion and a Random Forest classifier. This exceeds
the best pretrained neural network result of~73.8\%\ by~1.0 percentage points, and closes
the gap with the best handcrafted result of~77.8\%\ to 3.0 percentage points.
This suggests that training directly on malware images produces more
discriminative~512-dimensional embeddings than those transferred from
ImageNet pretraining, even at a fraction of the embedding dimensionality
used by VGG16.
Figure~\ref{fig:custom_cm} shows the confusion matrix for the best
custom CNN configuration (\grayscale\ + Random Forest) on the
test set. 


\subsubsection{Effect of Classifier}\label{sec:custom_classifier}

Unlike the pretrained neural network track, where Random Forest dominated uniformly,
the custom CNN track shows a more varied picture across classifiers.
SVM performs competitively across all conversion types in this track,
in contrast to its consistently poor performance in the pretrained CNN
track. This may be because the custom CNN embeddings are~512-dimensional
rather than up to~4,096-dimensional, making the feature space more
tractable for kernel methods.

\subsubsection{Effect of Conversion Type}\label{sec:custom_conversion}

The \grayscale\ conversion produces the strongest results, with all
five classifiers achieving between~73.8\%\ and~75.2\%\ accuracy. The \hit\ 
conversion is second-best with~74.8\%, and \entropyHilbert\ third with~73.1\%. 
A notable finding is that the bigram-based conversions, which produced
the weakest results in both the handcrafted and pretrained neural network tracks, perform
considerably better here: \bigramCartesian\ reaches~72.1\%\ and
\bigramPolar\ is at~71.6\%. This suggests that training a CNN from
scratch on these representations allows it to learn structural patterns that
are not captured by either handcrafted descriptors or ImageNet-pretrained
embeddings.

\subsubsection{Comparison with Previous Tracks}\label{sec:custom_vs_others}

Table~\ref{tab:track_summary} compares the best result from each feature-derivation track.
We observe that the custom CNN track outperforms the pretrained neural network track, 
despite producing a substantially more compact embedding. 
The handcrafted track is the strongest single-model approach.

\begin{table}[!htb]
\centering
\caption{Best individual result per feature track}\label{tab:track_summary}
\begin{adjustbox}{scale=0.725}
\begin{tabular}{llcc}
\toprule
\textbf{Track} & \textbf{Best configuration} & \textbf{Accuracy} & \textbf{F1} \\
\midrule
Handcrafted    & \hit\ + Random Forest (103 dims)  & 0.7782 & 0.7781 \\
Custom CNN     & \grayscale\ + Random Forest (512 dims)  & 0.7482 & 0.7480 \\
Pretrained neural network & ResNet50 / \grayscale\ + Random Forest (2,048 dims) & 0.7382 & 0.7378 \\
\bottomrule
\end{tabular}
\end{adjustbox}
\end{table}

\subsection{Soft Voting Ensemble}\label{sec:results_ensemble}

Our soft voting ensemble ranks all track configurations on the selection-validation
split and evaluates a range of pools, from single-track to fully cross-track
(Table~\ref{tab:ensemble_pools}). Each voter is trained
independently on the training split and produces a~17-dimensional probability
vector over the malware families. The final prediction is the class with the
highest average probability across the voters in a pool. Table~\ref{tab:pool_results}
shows the accuracy achieved by each voting pool configuration.

\begin{table}[!htb]
\centering
\caption{Soft voting ensemble results for all pool configurations}\label{tab:pool_results}
\begin{adjustbox}{scale=0.85}
\begin{tabular}{lcccc}
\toprule
\textbf{Pool} & \textbf{Voters} & \textbf{Val-select} & \textbf{Test} & \textbf{Test F1} \\
\midrule
Handcrafted track, top 15    & 15 & 0.7865 & \textbf{0.8018} & 0.8027 \\
Pretrained track, top 15     & 15 & 0.7482 & 0.7641 & 0.7642 \\
Custom CNN track, top 15     & 15 & 0.7624 & 0.7841 & 0.7836 \\
\midrule
Global top 3                 & \zz3 & 0.7600 & 0.7771 & 0.7780 \\
Global top 10                & 10 & 0.7618 & 0.7806 & 0.7812 \\
Global top 15                & 15 & 0.7541 & 0.7794 & 0.7790 \\
Global top 20                & 20 & 0.7571 & 0.7782 & 0.7779 \\
Global top 25                & 25 & 0.7629 & 0.7841 & 0.7835 \\
Global top 30                & 30 & 0.7618 & 0.7853 & 0.7849 \\
\midrule
Top 3 per track              & \zz9 & 0.7453 & 0.7747 & 0.7743 \\
Top 5 per track              & 15 & 0.7476 & 0.7747 & 0.7743 \\
Top 10 per track             & 30 & 0.7541 & 0.7806 & 0.7801 \\
Top 15 per track             & 45 & 0.7676 & 0.7894 & 0.7892 \\
\bottomrule
\end{tabular}
\end{adjustbox}
\end{table}


The best soft voting pool, the fifteen top-ranked handcrafted voters selected on
the selection-validation split, achieves~80.2\%\ accuracy on the test set, a gain
of~2.4 percentage points over the best individual model. Although modest, this
improvement is statistically significant by McNemar's test ($p<0.001$; see
Section~\ref{sec:statistics}). Because every voter and every pool is ranked on the
selection-validation split, with the test set reserved for a single final evaluation,
these numbers are free of the selection-induced optimism that arises when voters are
chosen on the test set. The cross-track pool of the fifteen best voters per track
reaches~78.9\%, and the best global top-$k$ pool reaches~78.5\%; no configuration
exceeds the handcrafted-only pool. This indicates that the handcrafted representations
dominate, while the pretrained and custom-CNN tracks contribute a small but
quantifiable complementary benefit, which we analyze in Section~\ref{sec:diversity}.

\subsubsection{Per-Class Performance}\label{sec:perclass}

Table~\ref{tab:perclass_summary} reports the F1-score and accuracy for
each malware family under the full~15-voter ensemble. 
\texttt{DCRat} is the most accurately classified family (F1~0.990), followed by
\texttt{Makoob} (F1~0.961). No family is classified perfectly. Note that a family
can have high per-family accuracy, meaning most of its own test samples are
correctly classified, yet a lower F1 if samples from other families are incorrectly
predicted as belonging to it, which reduces precision. \texttt{Seraph} illustrates
this: it has a recall of~0.980 but a precision of only~0.594, giving an F1 of~0.740.
The three most difficult families are \texttt{Agensla} (F1~0.688), \texttt{Stealerc}
(F1~0.708), and \texttt{Taskun} (F1~0.728), which together account for a large share
of the classification errors. The macro-averaged F1 is~0.803.

\begin{table}[!htb]
\centering
\caption{Per-class results for the 15-voter ensemble (100 samples per family)}\label{tab:perclass_summary}
\begin{adjustbox}{scale=0.85}
\begin{tabular}{lcc}
\toprule
\textbf{Family} & \textbf{F1} & \textbf{Accuracy} \\
\midrule
\texttt{DCRat}  & 0.9900 & 0.9900 \\
\texttt{Makoob}  & 0.9612 & 0.9900 \\
\texttt{SnakeLogger}  & 0.9053 & 0.8600 \\
\texttt{Crysan}  & 0.8691 & 0.8300 \\
\texttt{Strab}  & 0.8333 & 0.8500 \\
\texttt{Remcos}  & 0.8290 & 0.8000 \\
\texttt{Mokes}  & 0.8058 & 0.8300 \\
\texttt{Convagent}  & 0.7979 & 0.7500 \\
\texttt{Zenpak}  & 0.7822 & 0.7900 \\
\texttt{Androm}  & 0.7816 & 0.6800 \\
\texttt{Noon}  & 0.7556 & 0.6800 \\
\texttt{Injuke}  & 0.7437 & 0.7400 \\
\texttt{Seraph}  & 0.7396 & 0.9800 \\
\texttt{Crypt}  & 0.7283 & 0.6700 \\
\texttt{Taskun}  & 0.7281 & 0.7900 \\
\texttt{Stealerc}  & 0.7081 & 0.7400 \\
\texttt{Agensla}  & 0.6875 & 0.6600 \\
\midrule
Macro average & 0.8027 & 0.8018 \\
\bottomrule
\end{tabular}
\end{adjustbox}
\end{table}

\subsubsection{Confusion Analysis}\label{sec:confusion}

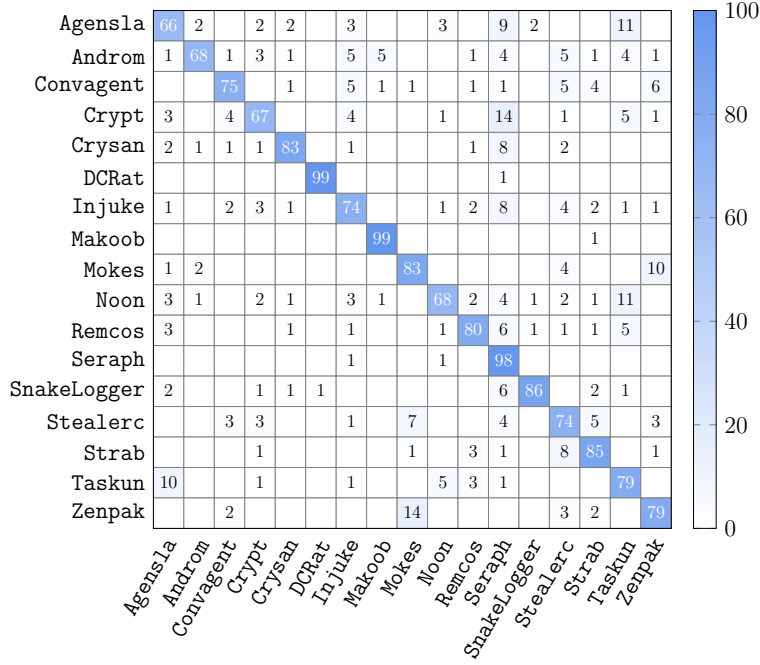
\begin{figure}[!htb]
\centering
\begin{tikzpicture}[scale=0.60]
    \begin{axis}[
        width=13cm,
        height=13cm,
	colormap={bluewhite}{color=(white) rgb255=(100,149,237)},
        xticklabels={
        Agensla,
        Androm,
        Convagent,
        Crypt,
        Crysan,
        DCRat,
        Injuke,
        Makoob,
        Mokes,
        Noon,
        Remcos,
        Seraph,
        SnakeLogger,
        Stealerc,
        Strab,
        Taskun,
        Zenpak
        },
        xtick={0,...,16},
        xtick style={draw=none},
	xticklabel style={scale=1.25,anchor=east,rotate=60,yshift=-5pt,font=\tt},
        yticklabels={
        Agensla,
        Androm,
        Convagent,
        Crypt,
        Crysan,
        DCRat,
        Injuke,
        Makoob,
        Mokes,
        Noon,
        Remcos,
        Seraph,
        SnakeLogger,
        Stealerc,
        Strab,
        Taskun,
        Zenpak
        },
        ytick={0,...,16},
        ytick style={draw=none},
        enlargelimits=false,
        yticklabel style={scale=1.25,font=\tt},
        colorbar,
        colorbar style={
            ytick={0,20,40,60,80,100},
            yticklabels={0,20,40,60,80,100},
            yticklabel={\pgfmathprintnumber\tick},
            yticklabel style={
            		scale=1.25,
            		/pgf/number format/fixed,
			/pgf/number format/fixed zerofill,
			/pgf/number format/precision=0}
        },
        point meta min=0,
        point meta max=100,
        nodes near coords={\pgfmathprintnumber\pgfplotspointmeta},
        nodes near coords black white/.style={
            small value/.style={
                yshift=-7pt,
                text=black,
                /pgf/number format/fixed,
                /pgf/number format/precision=0,
                /pgf/number format/zerofill=true,
                scale=0.95,
            },
            large value/.style={
                yshift=-7pt,
                text=white,
                /pgf/number format/fixed,
                /pgf/number format/precision=0,
                /pgf/number format/zerofill=true,
                scale=0.95,
            },
            every node near coord/.style={
                check for zero/.code={
                    \pgfmathfloatifflags{\pgfplotspointmeta}{0}{
                        \pgfkeys{/tikz/coordinate}
                    }{
                        \begingroup
                        \pgfkeys{/pgf/fpu}
                        \pgfmathparse{\pgfplotspointmeta<#1}
                        \global\let\result=\pgfmathresult
                        \endgroup
                        %
                        %
                        \pgfmathfloatcreate{1}{1.0}{0}
                        \let\ONE=\pgfmathresult
                        \ifx\result\ONE
                            \pgfkeysalso{/pgfplots/small value}
                        \else
                            \pgfkeysalso{/pgfplots/large value}
                        \fi
                    }
                },
                check for zero,
            },
        },
        nodes near coords black white=50,
    ]
\addplot[
  matrix plot,
  mesh/cols=17,
  point meta=explicit,
  draw=gray
] table [x=x, y=y, meta=C] {
x y C
0 0 66
1 0 2
2 0 0
3 0 2
4 0 2
5 0 0
6 0 3
7 0 0
8 0 0
9 0 3
10 0 0
11 0 9
12 0 2
13 0 0
14 0 0
15 0 11
16 0 0
0 1 1
1 1 68
2 1 1
3 1 3
4 1 1
5 1 0
6 1 5
7 1 5
8 1 0
9 1 0
10 1 1
11 1 4
12 1 0
13 1 5
14 1 1
15 1 4
16 1 1
0 2 0
1 2 0
2 2 75
3 2 0
4 2 1
5 2 0
6 2 5
7 2 1
8 2 1
9 2 0
10 2 1
11 2 1
12 2 0
13 2 5
14 2 4
15 2 0
16 2 6
0 3 3
1 3 0
2 3 4
3 3 67
4 3 0
5 3 0
6 3 4
7 3 0
8 3 0
9 3 1
10 3 0
11 3 14
12 3 0
13 3 1
14 3 0
15 3 5
16 3 1
0 4 2
1 4 1
2 4 1
3 4 1
4 4 83
5 4 0
6 4 1
7 4 0
8 4 0
9 4 0
10 4 1
11 4 8
12 4 0
13 4 2
14 4 0
15 4 0
16 4 0
0 5 0
1 5 0
2 5 0
3 5 0
4 5 0
5 5 99
6 5 0
7 5 0
8 5 0
9 5 0
10 5 0
11 5 1
12 5 0
13 5 0
14 5 0
15 5 0
16 5 0
0 6 1
1 6 0
2 6 2
3 6 3
4 6 1
5 6 0
6 6 74
7 6 0
8 6 0
9 6 1
10 6 2
11 6 8
12 6 0
13 6 4
14 6 2
15 6 1
16 6 1
0 7 0
1 7 0
2 7 0
3 7 0
4 7 0
5 7 0
6 7 0
7 7 99
8 7 0
9 7 0
10 7 0
11 7 0
12 7 0
13 7 0
14 7 1
15 7 0
16 7 0
0 8 1
1 8 2
2 8 0
3 8 0
4 8 0
5 8 0
6 8 0
7 8 0
8 8 83
9 8 0
10 8 0
11 8 0
12 8 0
13 8 4
14 8 0
15 8 0
16 8 10
0 9 3
1 9 1
2 9 0
3 9 2
4 9 1
5 9 0
6 9 3
7 9 1
8 9 0
9 9 68
10 9 2
11 9 4
12 9 1
13 9 2
14 9 1
15 9 11
16 9 0
0 10 3
1 10 0
2 10 0
3 10 0
4 10 1
5 10 0
6 10 1
7 10 0
8 10 0
9 10 1
10 10 80
11 10 6
12 10 1
13 10 1
14 10 1
15 10 5
16 10 0
0 11 0
1 11 0
2 11 0
3 11 0
4 11 0
5 11 0
6 11 1
7 11 0
8 11 0
9 11 1
10 11 0
11 11 98
12 11 0
13 11 0
14 11 0
15 11 0
16 11 0
0 12 2
1 12 0
2 12 0
3 12 1
4 12 1
5 12 1
6 12 0
7 12 0
8 12 0
9 12 0
10 12 0
11 12 6
12 12 86
13 12 0
14 12 2
15 12 1
16 12 0
0 13 0
1 13 0
2 13 3
3 13 3
4 13 0
5 13 0
6 13 1
7 13 0
8 13 7
9 13 0
10 13 0
11 13 4
12 13 0
13 13 74
14 13 5
15 13 0
16 13 3
0 14 0
1 14 0
2 14 0
3 14 1
4 14 0
5 14 0
6 14 0
7 14 0
8 14 1
9 14 0
10 14 3
11 14 1
12 14 0
13 14 8
14 14 85
15 14 0
16 14 1
0 15 10
1 15 0
2 15 0
3 15 1
4 15 0
5 15 0
6 15 1
7 15 0
8 15 0
9 15 5
10 15 3
11 15 1
12 15 0
13 15 0
14 15 0
15 15 79
16 15 0
0 16 0
1 16 0
2 16 2
3 16 0
4 16 0
5 16 0
6 16 0
7 16 0
8 16 14
9 16 0
10 16 0
11 16 0
12 16 0
13 16 3
14 16 2
15 16 0
16 16 79
};
\end{axis}
\end{tikzpicture}
\caption{Confusion matrix for the 15-voter soft voting ensemble}
\label{fig:app_confusion}
\end{figure}

Figure~\ref{fig:app_confusion} shows the~$17 \times 17$ confusion
matrix for the full~15-voter soft voting ensemble on the test set.
Rows represent true family labels and columns represent predicted
labels, while diagonal entries are correct classifications.


Table~\ref{tab:confused_pairs} lists the most frequently confused
family pairs under the full ensemble. The most prominent confusions are between \texttt{Mokes} and \texttt{Zenpak}, which
are confused in both directions (14 and~10 errors), and \texttt{Crypt} predicted as
\texttt{Seraph} (14 errors). \texttt{Taskun} acts as a frequent sink, with
\texttt{Noon} and \texttt{Agensla} each misclassified as \texttt{Taskun}. These
recurring errors suggest that the visual patterns of these families are relatively
similar across the eight conversion types.

\begin{table}[!htb]
\centering
\caption{Most confused family pairs (15-voter ensemble)}\label{tab:confused_pairs}
\begin{adjustbox}{scale=0.85}
\begin{tabular}{llc}
\toprule
\multicolumn{2}{c}{\hspace*{-0.125in}\textbf{Family}} & \multirow{2}{*}{\raisebox{-2pt}{\textbf{Count}}} \\ \cmidrule(lr){1-2}
\textbf{True} & \textbf{Predicted} \\
\midrule
\texttt{Zenpak}   & \texttt{Mokes} & 14 \\
\texttt{Crypt}   & \texttt{Seraph} & 14 \\
\texttt{Noon}   & \texttt{Taskun} & 11 \\
\texttt{Agensla}   & \texttt{Taskun} & 11 \\
\texttt{Taskun}   & \texttt{Agensla} & 10 \\
\texttt{Mokes}   & \texttt{Zenpak} & 10 \\
\texttt{Agensla}   & \texttt{Seraph} & \zz9 \\
\texttt{Strab}   & \texttt{Stealerc} & \zz8 \\
\texttt{Injuke}   & \texttt{Seraph} & \zz8 \\
\texttt{Crysan}   & \texttt{Seraph} & \zz8 \\
\texttt{Stealerc}   & \texttt{Mokes} & \zz7 \\
\texttt{SnakeLogger}   & \texttt{Seraph} & \zz6 \\
\texttt{Remcos}   & \texttt{Seraph} & \zz6 \\
\bottomrule
\end{tabular}
\end{adjustbox}
\end{table}

\subsection{Ensemble Diversity Analysis}\label{sec:diversity}

To explain why the soft voting ensemble improves only modestly over the best
individual model, and to substantiate the claim that the three feature tracks provide
complementary information, we quantify voter diversity using standard pairwise
measures. For a representative pool of the five best voters per track (fifteen voters),
Table~\ref{tab:diversity} reports the mean pairwise disagreement, Cohen's~$\kappa$, and
the Yule Q-statistic, evaluated on the test predictions and separated into
within-track and cross-track voter pairs.

\begin{table}[!htb]
\centering
\caption{Ensemble diversity: mean pairwise statistics within vs.\ across feature tracks 
(higher disagreement / lower $\kappa$ and $Q$ mean more diverse)}\label{tab:diversity}
\begin{adjustbox}{scale=0.85}
\begin{tabular}{lccc}
\toprule
\textbf{Voter pairs} & \textbf{Disagreement} & \textbf{Cohen's $\kappa$} & \textbf{Q-statistic} \\
\midrule
Within the same track & 0.165 & 0.825 & 0.964 \\
Across different tracks & 0.215 & 0.771 & 0.945 \\
\bottomrule
\end{tabular}
\end{adjustbox}
\end{table}

\begin{figure}[!htb]
\centering
\begin{tikzpicture}[scale=0.80]
    \begin{axis}[
        width=11.5cm,
        height=11.5cm,
	colormap={redwhite}{color=(white) color=(red)},
        xticklabels={
\textrm{HC:HIT+RF},
\textrm{HC:Grayscale+RF},
\textrm{HC:Byteclass Hilbert+XGB},
\textrm{HC:Grayscale+XGB},
\textrm{HC:HIT+XGB},
\textrm{PT:VGG16/Grayscale+RF},
\textrm{PT:ResNet50/Grayscale+RF},
\textrm{PT:ResNet50/Grayscale+XGB},
\textrm{PT:ResNet50/Byteclass+XGB},
\textrm{PT:VGG16/Byteclass+XGB},
\textrm{CNN:Grayscale+RF},
\textrm{CNN:HIT+SVM},
\textrm{CNN:HIT+XGB},
\textrm{CNN:HIT+CB},
\textrm{CNN:HIT+RF}
        },
        xtick={0,...,14},
        xtick style={draw=none},
	xticklabel style={anchor=east,rotate=60,yshift=-5pt,scale=0.85},
        yticklabels={
\textrm{HC:HIT+RF},
\textrm{HC:Grayscale+RF},
\textrm{HC:Byteclass Hilbert+XGB},
\textrm{HC:Grayscale+XGB},
\textrm{HC:HIT+XGB},
\textrm{PT:VGG16/Grayscale+RF},
\textrm{PT:ResNet50/Grayscale+RF},
\textrm{PT:ResNet50/Grayscale+XGB},
\textrm{PT:ResNet50/Byteclass+XGB},
\textrm{PT:VGG16/Byteclass+XGB},
\textrm{CNN:Grayscale+RF},
\textrm{CNN:HIT+SVM},
\textrm{CNN:HIT+XGB},
\textrm{CNN:HIT+CB},
\textrm{CNN:HIT+RF}
        },
        ytick={0,...,14},
        ytick style={draw=none},
        enlargelimits=false,
        yticklabel style={scale=0.85},
        colorbar,
        colorbar style={
            ytick={0.6,0.7,0.8,0.9,1.0},
            yticklabels={0.6,0.7,0.8,0.9,1.0},
            yticklabel={\pgfmathprintnumber\tick},
            yticklabel style={
            		scale=0.9,
  			/pgf/number format/.cd,
  			fixed,
  			fixed zerofill,
  			precision=1}
        },
        point meta min=0.6,
        point meta max=1.0,
        nodes near coords={\pgfmathprintnumber\pgfplotspointmeta},
        nodes near coords black white/.style={
            small value/.style={
                yshift=-5.0pt,
                text=black,
                /pgf/number format/fixed,
                /pgf/number format/precision=3,
                /pgf/number format/zerofill=true,
                scale=0.6,
            },
            large value/.style={
                yshift=-5.0pt,
                text=white,
                /pgf/number format/fixed,
                /pgf/number format/precision=3,
                /pgf/number format/zerofill=true,
                scale=0.6,
            },
            every node near coord/.style={
                check for zero/.code={
                    \pgfmathfloatifflags{\pgfplotspointmeta}{0}{
                        \pgfkeys{/tikz/coordinate}
                    }{
                        \begingroup
                        \pgfkeys{/pgf/fpu}
                        \pgfmathparse{\pgfplotspointmeta<#1}
                        \global\let\result=\pgfmathresult
                        \endgroup
                        %
                        %
                        \pgfmathfloatcreate{1}{1.0}{0}
                        \let\ONE=\pgfmathresult
                        \ifx\result\ONE
                            \pgfkeysalso{/pgfplots/small value}
                        \else
                            \pgfkeysalso{/pgfplots/large value}
                        \fi
                    }
                },
                check for zero,
            },
        },
        nodes near coords black white=0.8,
    ]
        \addplot[
            matrix plot,
            mesh/cols=15,
            point meta=explicit,draw=gray
        ] table [meta=C] {
            x y C
0	0	1
1	0	0.7454
2	0	0.7494
3	0	0.6967
4	0	0.8228
5	0	0.7248
6	0	0.7032
7	0	0.686
8	0	0.6749
9	0	0.6748
10	0	0.6691
11	0	0.6823
12	0	0.6917
13	0	0.6812
14	0	0.6914
0	1	0.7454
1	1	1
2	1	0.6812
3	1	0.8322
4	1	0.6796
5	1	0.7314
6	1	0.7389
7	1	0.7114
8	1	0.6845
9	1	0.6752
10	1	0.6893
11	1	0.6571
12	1	0.6728
13	1	0.6818
14	1	0.6823
0	2	0.7494
1	2	0.6812
2	2	1
3	2	0.655
4	2	0.7422
5	2	0.6797
6	2	0.6996
7	2	0.6744
8	2	0.6995
9	2	0.668
10	2	0.6966
11	2	0.6705
12	2	0.6831
13	2	0.67
14	2	0.6739
0	3	0.6967
1	3	0.8322
2	3	0.655
3	3	1
4	3	0.6646
5	3	0.7021
6	3	0.6816
7	3	0.6631
8	3	0.6519
9	3	0.6594
10	3	0.6813
11	3	0.6523
12	3	0.6616
13	3	0.6673
14	3	0.6711
0	4	0.8228
1	4	0.6796
2	4	0.7422
3	4	0.6646
4	4	1
5	4	0.6789
6	4	0.6863
7	4	0.6344
8	4	0.6574
9	4	0.6454
10	4	0.6564
11	4	0.669
12	4	0.6816
13	4	0.6713
14	4	0.6815
0	5	0.7248
1	5	0.7314
2	5	0.6797
3	5	0.7021
4	5	0.6789
5	5	1
6	5	0.7636
7	5	0.7121
8	5	0.7005
9	5	0.6802
10	5	0.686
11	5	0.6566
12	5	0.6753
13	5	0.6749
14	5	0.6789
0	6	0.7032
1	6	0.7389
2	6	0.6996
3	6	0.6816
4	6	0.6863
5	6	0.7636
6	6	1
7	6	0.7829
8	6	0.6996
9	6	0.682
10	6	0.6874
11	6	0.6734
12	6	0.6799
13	6	0.6825
14	6	0.6834
0	7	0.686
1	7	0.7114
2	7	0.6744
3	7	0.6631
4	7	0.6344
5	7	0.7121
6	7	0.7829
7	7	1
8	7	0.6778
9	7	0.6674
10	7	0.6288
11	7	0.6388
12	7	0.6273
13	7	0.639
14	7	0.6218
0	8	0.6749
1	8	0.6845
2	8	0.6995
3	8	0.6519
4	8	0.6574
5	8	0.7005
6	8	0.6996
7	8	0.6778
8	8	1
9	8	0.7119
10	8	0.6418
11	8	0.6812
12	8	0.673
13	8	0.6757
14	8	0.6768
0	9	0.6748
1	9	0.6752
2	9	0.668
3	9	0.6594
4	9	0.6454
5	9	0.6802
6	9	0.682
7	9	0.6674
8	9	0.7119
9	9	1
10	9	0.6131
11	9	0.6721
12	9	0.6392
13	9	0.6631
14	9	0.6641
0	10	0.6691
1	10	0.6893
2	10	0.6966
3	10	0.6813
4	10	0.6564
5	10	0.686
6	10	0.6874
7	10	0.6288
8	10	0.6418
9	10	0.6131
10	10	1
11	10	0.6771
12	10	0.6804
13	10	0.6643
14	10	0.6776
0	11	0.6823
1	11	0.6571
2	11	0.6705
3	11	0.6523
4	11	0.669
5	11	0.6566
6	11	0.6734
7	11	0.6388
8	11	0.6812
9	11	0.6721
10	11	0.6771
11	11	1
12	11	0.8756
13	11	0.8846
14	11	0.8829
0	12	0.6917
1	12	0.6728
2	12	0.6831
3	12	0.6616
4	12	0.6816
5	12	0.6753
6	12	0.6799
7	12	0.6273
8	12	0.673
9	12	0.6392
10	12	0.6804
11	12	0.8756
12	12	1
13	12	0.8857
14	12	0.9211
0	13	0.6812
1	13	0.6818
2	13	0.67
3	13	0.6673
4	13	0.6713
5	13	0.6749
6	13	0.6825
7	13	0.639
8	13	0.6757
9	13	0.6631
10	13	0.6643
11	13	0.8846
12	13	0.8857
13	13	1
14	13	0.924
0	14	0.6914
1	14	0.6823
2	14	0.6739
3	14	0.6711
4	14	0.6815
5	14	0.6789
6	14	0.6834
7	14	0.6218
8	14	0.6768
9	14	0.6641
10	14	0.6776
11	14	0.8829
12	14	0.9211
13	14	0.924
14	14	1
         };
         \draw[black, line width=1.5pt] (-0.5,4.5) -- (14.5,4.5);
         \draw[black, line width=1.5pt] (-0.5,9.5) -- (14.5,9.5);
         \draw[black, line width=1.5pt] (4.5,-0.5) -- (4.5,14.5);
         \draw[black, line width=1.5pt] (9.5,-0.5) -- (9.5,14.5);
    \end{axis}
\end{tikzpicture}
\caption{Per-sample error correlation between voters (five best per track, where
\textrm{HC} is Handcrafted track,
\textrm{PT} is Pretrained track, 
\textrm{CNN} is Custom CNN track,
\textrm{RF} is Random Forest,
\textrm{XGB} is XGBoost, and
\textrm{CB} is CatBoost)
}
\label{fig:errcorr}
\end{figure}
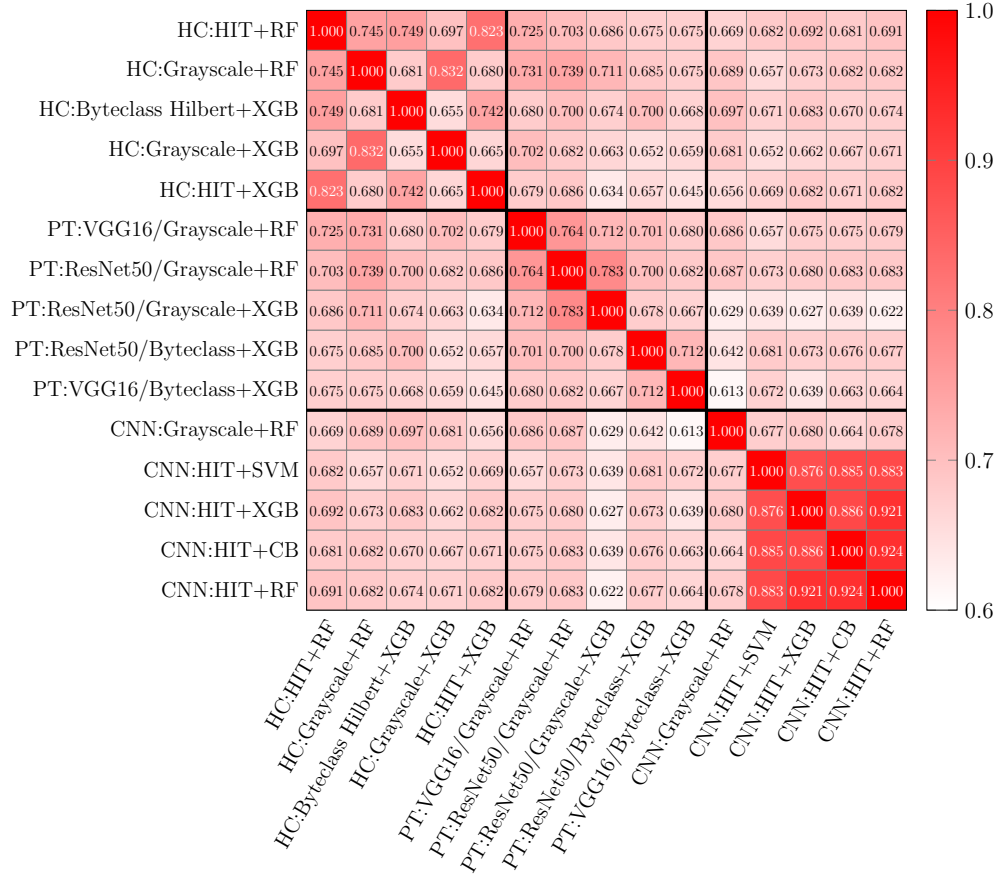

Voters within the same track are highly correlated (mean~$\kappa = 0.83$, with a
disagreement of only~0.17), confirming that they are largely redundant. Cross-track
voter pairs are more diverse (mean~$\kappa = 0.77$, disagreement~0.22), which confirms
that the three representations are genuinely complementary. The increase in diversity
is, however, modest, and the pretrained and custom-CNN voters are individually weaker
than the handcrafted voters; consequently, equal-weight probability averaging cannot
translate the additional diversity into a large accuracy gain. Figure~\ref{fig:errcorr}
shows the per-sample error-correlation matrix, in which the three within-track blocks
of higher correlation are clearly visible.


\begin{figure}[!htb]
\centering
\includegraphics[scale=0.5]{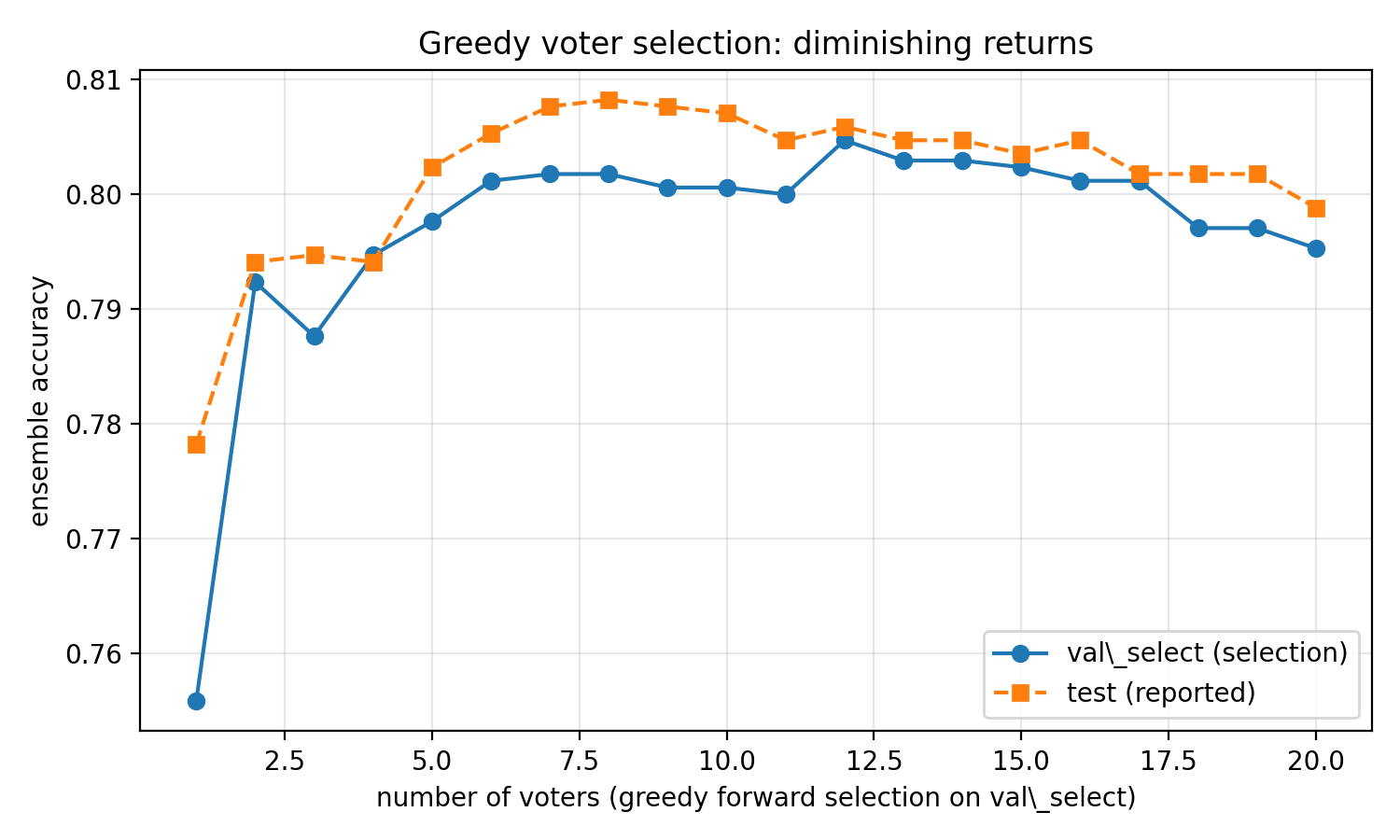}
\caption{Greedy forward selection of voters (ranked on the selection-validation split)}\label{fig:greedy}
\end{figure}

A leave-one-out ablation of the best (handcrafted) pool, reported in
Table~\ref{tab:ablation}, shows that removing any single voter changes test accuracy by
at most a fraction of a percentage point (largest effect~$+0.88$, from dropping the
weakest voter), consistent with the high within-track redundancy. Finally,
Figure~\ref{fig:greedy} traces a greedy forward selection of voters ranked on the
selection-validation split: accuracy rises quickly over the first five voters and then
fluctuates narrowly, between roughly~79\%\ and~81\%\ on the test set, up to twenty
voters, with no further meaningful gain, corroborating that approximately~80\%\
accuracy is close to the practical ceiling for this soft voting approach on this
dataset.

\begin{table}[!htb]
\centering
\caption{Leave-one-out voter contribution in the headline pool 
($\Delta$ test accuracy when the voter is removed)}\label{tab:ablation}
\begin{adjustbox}{scale=0.85}
\begin{tabular}{lc}
\toprule
\multirow{2}{*}{\textbf{Removed voter}} & \textbf{$\bm{\Delta}$ accuracy} \\
     & \textbf{(percentage)} \\
\midrule
$\mbox{\spiral}+\mbox{XGBoost}$ & $\hbox{}+0.88$ \\
$\mbox{\spiral}+\mbox{Random Forest}$ & $\hbox{}+0.53$ \\
$\mbox{\bigramCartesian}+\mbox{XGBoost}$ & $\hbox{}+0.53$ \\
$\mbox{\bigramCartesian}+\mbox{Random Forest}$ & $\hbox{}+0.41$ \\
$\mbox{\bigramPolar}+\mbox{Random Forest}$ & $\hbox{}+0.41$ \\
$\mbox{\grayscale}+\mbox{XGBoost}$ & $\hbox{}+0.12$ \\
$\mbox{\byteclass}+\mbox{Random Forest}$ & $\hbox{}+0.06$ \\
$\mbox{\hit}+\mbox{XGBoost}$ & $\hbox{}+0.00$ \\
$\mbox{\grayscale}+\mbox{Random Forest}$ & $\hbox{}-0.06$\kern 1pt \\
$\mbox{\byteclass}+\mbox{XGBoost}$ & $\hbox{}-0.06$\kern 1pt \\
$\mbox{\hit}+\mbox{Random Forest}$ & $\hbox{}-0.12$\kern 1pt \\
$\mbox{\byteclassHilbert}+\mbox{XGBoost}$ & $\hbox{}-0.18$\kern 1pt \\
$\mbox{\entropyHilbert}+\mbox{Random Forest}$ & $\hbox{}-0.18$\kern 1pt \\
$\mbox{\byteclassHilbert}+\mbox{Random Forest}$ & $\hbox{}-0.18$\kern 1pt \\
$\mbox{\entropyHilbert}+\mbox{XGBoost}$ & $\hbox{}-0.24$\kern 1pt \\
\bottomrule
\end{tabular}
\end{adjustbox}
\end{table}

\subsection{Statistical Validation}\label{sec:statistics}

Because our conclusions rest on a single training--validation--test split, we assess
the robustness of the reported accuracies with bootstrap confidence intervals and a
paired significance test. Table~\ref{tab:ci} reports test accuracy together with~95\%\
bootstrap confidence intervals (10{,}000 resamples of the~1{,}700 test predictions) for
the headline ensemble, the best cross-track ensemble, and the best voter of each track.

\begin{table}[!htb]
\centering
\caption{Test accuracy with bootstrap 95\% confidence intervals (10{,}000 resamples, $n=1{,}700$)}\label{tab:ci}
\begin{adjustbox}{scale=0.785}
\begin{tabular}{lcc}
\toprule
\multirow{2}{*}{\textbf{Configuration}} & \multirow{2}{*}{\textbf{Accuracy}} & \textbf{Confidence} \\
   &  & \textbf{interval} \\
\midrule
Headline ensemble (handcrafted top-15) & 0.8018 & [0.7824, 0.8206] \\
Cross-track ensemble (top-15 per track) & 0.7894 & [0.7700, 0.8088] \\
Best single voter (HIT + Random Forest ) & 0.7782 & [0.7576, 0.7976] \\
Best pretrained voter (ResNet50/Grayscale + Random Forest ) & 0.7382 & [0.7171, 0.7594] \\
Best custom-CNN voter (Grayscale + Random Forest ) & 0.7482 & [0.7271, 0.7688] \\
\bottomrule
\end{tabular}
\end{adjustbox}
\end{table}

The confidence intervals overlap across the strongest configurations, which is expected
given the size of the test set; a paired comparison is therefore more informative.
Comparing the headline ensemble against the best individual voter (HIT with a Random
Forest classifier) using McNemar's exact test yields~$p < 0.001$ (discordant
pairs~$b = 77$ and~$c = 37$). Hence the~2.4 percentage-point improvement of the
ensemble over the best individual model, although small, is statistically significant.
Repeating the deep-network tracks over multiple random seeds is computationally
expensive, as each custom-CNN training requires several hours on our hardware, so the
bootstrap and McNemar analyses provide split-level uncertainty estimates without that
cost.

\subsection{Discussion}\label{sec:results_summary}

Table~\ref{tab:final_summary} summarizes the key accuracy results across
all experiments conducted in this research.
Individual models from each track achieve between~73.8\%\ and~77.8\%\ accuracy.
Single-track ensembles of the top fifteen voters reach the~76\%\ to~80\%\ range.
Combining all three tracks does not exceed the best single-track pool: the strongest
ensemble reaches~80.2\%\ accuracy, a small but statistically significant gain over the
best individual model. The handcrafted representations dominate; the complementarity
of the three feature representations, where each track tends to make somewhat
different errors, yields only a modest benefit under equal-weight probability
averaging, as
quantified in Section~\ref{sec:diversity}. Notably, the ten highest-ranked individual configurations overall 
belong to the handcrafted track (Appendix~C), each outperforming the best model from either the 
pretrained or the custom-CNN track. 

\begin{table}[!htb]
\centering
\caption{Summary of best results}\label{tab:final_summary}
\begin{adjustbox}{scale=0.75}
\begin{tabular}{llc}
\toprule
\textbf{Method} & \textbf{Configuration} & \textbf{Accuracy} \\
\midrule
Previous benchmark~\cite{Agrawal_2025} & \grayscale\ + XGBoost (HOG) & 0.7512 \\
Handcrafted (best model)    & \hit\ + Random Forest  (103 features)          & 0.7782 \\
Pretrained CNN (best model) & ResNet50 / \grayscale\ + Random Forest  (2{,}048 dims) & 0.7382 \\
Custom CNN (best model)     & \grayscale\ + Random Forest  (512 dims)         & 0.7482 \\
Pretrained ensemble & 15 voters, pretrained only           & 0.7641 \\
Custom ensemble     & 15 voters, custom only               & 0.7841 \\
Cross-track ensemble & 15 voters per track (45 total)      & 0.7894 \\
Soft voting ensemble & Handcrafted, 15 voters & \textbf{0.8018} \\
\bottomrule
\end{tabular}
\end{adjustbox}
\end{table}

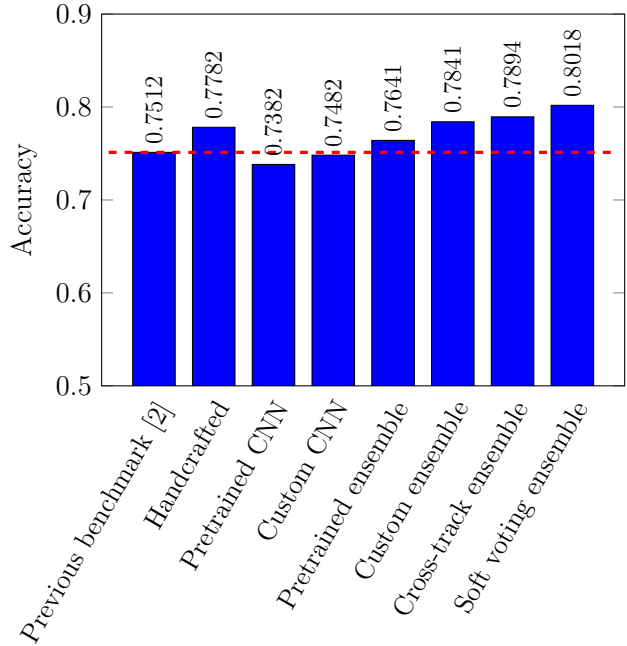
\begin{figure}[!htb]
\centering
\begin{tikzpicture}[scale=1.0, every node/.style={scale=1.0}]
\pgfkeys{/pgf/number format/.cd,1000 sep={}}
\begin{axis}[
        width  = 8.5cm,
        height = 6.5cm,
        ymin=0.50,ymax=0.9,
        ytick={0.5, 0.6, 0.7, 0.8, 0.9},
        major x tick style = transparent,
        ybar=5*\pgflinewidth,
        bar width=16.0pt,
        ylabel = {Accuracy},
        ylabel style = {scale = 0.9},
        symbolic x coords={A, B, C, D, E, F, G, H},
        xticklabels={Previous benchmark~\cite{Agrawal_2025}, 
        Handcrafted,
        Pretrained CNN,
        Custom CNN,
        Pretrained ensemble,
        Custom ensemble,
        Cross-track ensemble,
        Soft voting ensemble},
	y tick label style={scale=0.9,
    		/pgf/number format/.cd,
   		fixed,
   		fixed zerofill,
    		precision=1},
        xtick = data,
        x tick label style={scale=0.85,
	        	rotate=60,
		anchor=north east,
		inner sep=0mm
		},
        nodes near coords,
        every node near coord/.append style={rotate=90, scale=0.785,
        								   anchor=west, 
								   /pgf/number format/.cd,
								   fixed,
								   fixed zerofill,
								   precision=4},
        enlarge x limits=0.125,
        legend cell align=left,
        legend pos=south east,
]
\addplot [fill=blue,opacity=1.00]
coordinates {
(A, 0.7512)
(B, 0.7782)
(C, 0.7382)
(D, 0.7482)
(E, 0.7641)
(F, 0.7841)
(G, 0.7894)
(H, 0.8018)
};
\addplot[red,very thick,dashed,line legend,sharp plot,nodes near coords={},
    update limits=false,shorten >=-6mm,shorten <=-6mm] 
    coordinates {(A,0.7512) (H,0.7512)};
\end{axis}
\end{tikzpicture}
\caption{Bar graph of best results (dashed line corresponds to previous benchmark)}\label{bar:best}
\end{figure}

Finally, the results in Table~\ref{tab:final_summary} are given in the form of a bar graph
in Figure~\ref{bar:best}. This bar graph serves to emphasize that five of our seven models
listed in Table~\ref{tab:final_summary} exceed the previous benchmark result 
from~\cite{Agrawal_2025}.

\section{Conclusion and Future Work}\label{sec:conclusion}

In this chapter, we developed and analyzed a multi-track feature engineering and ensemble
classification system for malware family classification, using image-based
representations of binary executables. Three feature extraction tracks
were evaluated across eight malware-to-image conversion types, which were then combined
via a soft voting ensemble.

The handcrafted feature track achieved a best accuracy of~77.8\%\ using~103 features
selected from a~6,135-dimensional fused feature vector, outperforming the
HOG-only baseline of Agrawal et al.~\cite{Agrawal_2025}. The pretrained
neural network track achieved~73.8\%\ accuracy using ResNet50 embeddings. The custom CNN track
achieved a~74.8\%\ accuracy using~512-dimensional embeddings trained directly on
malware images, exceeding the pretrained track despite a
more compact representation.

The best soft voting pool, the fifteen top-ranked handcrafted voters selected on a
dedicated validation split, achieved~80.2\% accuracy on the test set of~1,700 samples
across~17 malware families, a statistically significant gain of~2.4 percentage points
over the best individual model (McNemar's test, $p<0.001$). Pool ablation experiments
showed that the single-track ensembles of the non-CNN, pretrained neural network, and
custom CNN pools achieve~80.2\%, 76.4\%, and~78.4\%, respectively; no cross-track pool
exceeds the handcrafted-only pool. Once test-set leakage is removed, the handcrafted
representations dominate, and the three feature representations are complementary only
to a modest degree, as quantified by our diversity analysis.



There are several directions in which this research can be extended. Since
the current ensemble averages the~15 voters' probability vectors with equal weight, weighting each output
by validation accuracy or applying more sophisticated combination strategies
such as stacking~\cite{StampEnsemble2021,Wolpert_1992} may further improve performance.
Fine-tuning the pretrained neural network models end-to-end on the malware images,
rather than using them as fixed feature extractors, is another natural extension.

The current dataset is limited to~17 families with~1,000 samples each. Experiments
on a larger and more diverse corpus would strengthen confidence in the generality of the
results. The confusion between families 
such as \texttt{Mokes}--\texttt{Zenpak} and \texttt{Injuke}--\texttt{Convagent}
suggests shared structural properties that may warrant binary-level analysis.

Finally, interpretability is an important direction for future work. Techniques
such as Grad-CAM~\cite{selvaraju2017gradcam}, 
HiResCAM~\cite{hirescam2020}, or SHAP~\cite{grad_cam} 
could identify which image regions and features
contribute most to predictions, and may also shed light on the observed
complementarity between the three feature types.

\bibliographystyle{plain}
\bibliography{references, references_vibha}



\section*{Appendix A}\label{app:a}

\titleformat{\section}{\normalfont\large\bfseries}{}{0em}{#1\ \thesection}
\setcounter{section}{0}
\renewcommand{\thesection}{\Alph{section}}
\renewcommand{\thesubsection}{A.\arabic{subsection}}
\setcounter{table}{0}
\renewcommand{\thetable}{A.\arabic{table}}
\setcounter{figure}{0}
\renewcommand{\thefigure}{A.\arabic{figure}}

This appendix contains supplementary tables for the handcrafted feature
extraction track. Section~\ref{app:fused_full} presents the complete fused
and reduced feature results. Section~\ref{app:composition} provides the
feature-type composition of each selected subset. Section~\ref{app:best_subset}
lists the individual features selected for the best performing model.

\subsection{Full Fused and Reduced Feature Results}\label{app:fused_full}

Table~\ref{tab:fused_full_table} contains the complete fused-and-reduced
results for all eight conversion types evaluated with Random Forest and XGBoost.

\begin{table}[!htb]
\centering
\caption{Complete fused and reduced feature results on the test set}\label{tab:fused_full_table}
\begin{adjustbox}{scale=0.85}
\begin{tabular}{llccccr}
\toprule
\textbf{Conversion} & \textbf{Classifier} &
\textbf{Accuracy} & \textbf{Precision} & \textbf{Recall} &
\textbf{F1} & \textbf{Subset} \\
\midrule
\byteclassHilbert\  & Random Forest      & 0.7794 & 0.7903 & 0.7794 & 0.7802 & 83 \\
\byteclassHilbert\  & XGBoost & 0.7771 & 0.7844 & 0.7771 & 0.7775 & 83 \\
\grayscale          & Random Forest      & 0.7753 & 0.7815 & 0.7753 & 0.7755 & 50 \\
\grayscale          & XGBoost & 0.7594 & 0.7626 & 0.7594 & 0.7596 & 50 \\
\hit                & Random Forest      & 0.7688 & 0.7863 & 0.7688 & 0.7703 & 160 \\
\hit                & XGBoost & 0.7659 & 0.7734 & 0.7659 & 0.7659 & 160 \\
\entropyHilbert            & Random Forest      & 0.7612 & 0.7725 & 0.7612 & 0.7618 & 67 \\
\entropyHilbert            & XGBoost & 0.7535 & 0.7606 & 0.7535 & 0.7538 & 67 \\
\byteclass          & Random Forest      & 0.7588 & 0.7657 & 0.7588 & 0.7591 & 83 \\
\byteclass          & XGBoost & 0.7465 & 0.7497 & 0.7465 & 0.7463 & 83 \\
\spiral             & Random Forest      & 0.7088 & 0.7258 & 0.7088 & 0.7082 & 388 \\
\spiral             & XGBoost & 0.7000 & 0.7037 & 0.7000 & 0.6982 & 388 \\
\bigramCartesian  & XGBoost & 0.7059 & 0.7104 & 0.7059 & 0.7049 & 484 \\
\bigramCartesian  & Random Forest      & 0.6918 & 0.7099 & 0.6918 & 0.6902 & 484 \\
\bigramPolar      & Random Forest      & 0.6918 & 0.7174 & 0.6918 & 0.6919 & 1{,}178 \\
\bigramPolar      & XGBoost & 0.6835 & 0.6932 & 0.6835 & 0.6828 & 1{,}178 \\
\bottomrule
\end{tabular}
\end{adjustbox}
\end{table}

\subsection{RFE Subset Composition}\label{app:composition}

Table~\ref{tab:app_composition} provides the number of HOG, Haralick, and
ADV38 features retained after two-stage RFE reduction for each conversion
type. These results support the discussion in Section~\ref{sec:results_noncnn}.

\begin{table}[!htb]
\centering
\caption{RFE-selected feature subset composition per conversion type}\label{tab:app_composition}
\begin{adjustbox}{scale=0.85}
\begin{tabular}{lrrcc}
\toprule
\textbf{Conversion} & \textbf{Total} & \textbf{HOG} &
\textbf{Haralick} & \textbf{ADV38} \\
\midrule
\byteclassHilbert\  &    83 &    62 & 11 & 10 \\
\grayscale          &    50 &    26 &  \zz9 & 15 \\
\hit                &   160 &   118 & 13 & 29 \\
\entropyHilbert            &    67 &    38 & 11 & 18 \\
\byteclass          &    83 &    53 & 13 & 17 \\
\spiral             &   388 &   352 & 13 & 23 \\
\bigramCartesian  &   484 &   449 & 13 & 22 \\
\bigramPolar      & 1{,}178 & 1{,}141 & 13 & 24 \\
\bottomrule
\end{tabular}
\end{adjustbox}
\end{table}

\subsection{Selected Feature Subset for Best Fused Model}\label{app:best_subset}

The best-performing fused model was obtained for the \byteclassHilbert\ 
conversion using Random Forest, achieving~0.7794 accuracy with~83 selected features.
Table~\ref{tab:best_feature_list} lists all selected feature indices and their
corresponding names, grouped by descriptor type.

\begin{table}[!htb]
\centering
\caption{Selected features for the best fused model (\byteclassHilbert, Random Forest)}\label{tab:best_feature_list}
\begin{adjustbox}{scale=0.725}
\begin{tabular}{cl|cl|cl|cl}
\toprule
\textbf{Index} & \textbf{Name} & \textbf{Index} & \textbf{Name} & \textbf{Index} & \textbf{Name} & \textbf{Index} & \textbf{Name} \\
\midrule
\zz0 & hog\_468 & 21 & hog\_4279 & 42 & hog\_1635 & 63 & haralick\_imc2 \\
\zz1 & hog\_1 & 22 & hog\_3843 & 43 & hog\_2 & 64 & haralick\_imc1 \\
\zz2 & hog\_3 & 23 & hog\_490 & 44 & hog\_234 & 65 & haralick\_idm \\
\zz3 & hog\_4 & 24 & hog\_474 & 45 & hog\_1440 & 66 & haralick\_diff\_var \\
\zz4 & hog\_3825 & 25 & hog\_961 & 46 & hog\_508 & 67 & haralick\_contrast \\
\zz5 & hog\_938 & 26 & hog\_38 & 47 & hog\_1170 & 68 & haralick\_sum\_var \\
\zz6 & hog\_3847 & 27 & hog\_11 & 48 & hog\_4158 & 69 & haralick\_diff\_entropy \\
\zz7 & hog\_1411 & 28 & hog\_1451 & 49 & hog\_3712 & 70 & haralick\_asm \\
\zz8 & hog\_489 & 29 & hog\_3280 & 50 & hog\_949 & 71 & haralick\_sum\_entropy \\
\zz9 & hog\_24 & 30 & hog\_3721 & 51 & hog\_2803 & 72 & haralick\_entropy \\
10 & hog\_956 & 31 & hog\_13 & 52 & hog\_5602 & 73 & adv\_edges\_count \\
11 & hog\_71 & 32 & hog\_947 & 53 & hog\_940 & 74 & adv\_entr\_hsv \\
12 & hog\_4284 & 33 & hog\_3829 & 54 & hog\_60 & 75 & adv\_avg\_edge\_length \\
13 & hog\_470 & 34 & hog\_1406 & 55 & hog\_5980 & 76 & adv\_snr \\
14 & hog\_504 & 35 & hog\_4941 & 56 & hog\_0 & 77 & adv\_hog\_entropy \\
15 & hog\_5112 & 36 & hog\_621 & 57 & hog\_22 & 78 & adv\_entr\_noise \\
16 & hog\_488 & 37 & hog\_486 & 58 & hog\_1019 & 79 & adv\_s\_mean \\
17 & hog\_472 & 38 & hog\_414 & 59 & hog\_1541 & 80 & adv\_lbp\_entropy \\
18 & hog\_3206 & 39 & hog\_44 & 60 & hog\_9 & 81 & adv\_entr\_color \\
19 & hog\_939 & 40 & hog\_5548 & 61 & hog\_2018 & 82 & adv\_g\_var \\
20 & hog\_18 & 41 & hog\_1638 & 62 & haralick\_correlation & 	---  & ---\\
\bottomrule
\end{tabular}
\end{adjustbox}
\end{table}

\section*{Appendix B}\label{app:b}

\titleformat{\section}{\normalfont\large\bfseries}{}{0em}{#1\ \thesection}
\setcounter{section}{0}
\renewcommand{\thesection}{\Alph{section}}
\renewcommand{\thesubsection}{B.\arabic{subsection}}
\setcounter{table}{0}
\renewcommand{\thetable}{B.\arabic{table}}
\setcounter{figure}{0}
\renewcommand{\thefigure}{B.\arabic{figure}}

This appendix contains the complete experimental results for the pretrained
CNN and custom CNN feature extraction tracks.
Section~\ref{app:pretrained} provides all~144 pretrained CNN results.
Section~\ref{app:custom} provides all~40 custom CNN results.
Section~\ref{app:ensemble_perclass} provides the full per-class classification
report for the 15-voter ensemble.

\subsection{Full Pretrained CNN Results}\label{app:pretrained}

Table~\ref{tab:app_pretrained} contains the complete results for all~144
pretrained CNN experiments across three CNN models, eight conversion types,
and six classifiers, sorted by accuracy in descending order.

{\centering
\scriptsize
\begin{longtable}{lllcccc}
\caption{Complete pretrained CNN results on the test set (144 experiments),
sorted by accuracy}\label{tab:app_pretrained} \\
\hline\hline
\textbf{CNN} & \textbf{Conversion} & \textbf{Classifier} &
\textbf{Accuracy} & \textbf{Precision} & \textbf{Recall} & \textbf{F1} \\
\hline
\endfirsthead
\hline
\textbf{CNN} & \textbf{Conversion} & \textbf{Classifier} &
\textbf{Accuracy} & \textbf{Precision} & \textbf{Recall} & \textbf{F1} \\
\hline
\endhead
VGG16    & \grayscale          & RF      & 0.7394 & 0.7507 & 0.7394 & 0.7390 \\
ResNet50 & \grayscale          & RF      & 0.7376 & 0.7490 & 0.7376 & 0.7377 \\
VGG16    & \byteclass          & RF      & 0.7376 & 0.7512 & 0.7376 & 0.7383 \\
ResNet50 & \byteclass          & RF      & 0.7329 & 0.7434 & 0.7329 & 0.7338 \\
ViT      & \byteclass          & RF      & 0.7271 & 0.7364 & 0.7271 & 0.7282 \\
ResNet50 & \grayscale          & XGB & 0.7247 & 0.7348 & 0.7247 & 0.7249 \\
ResNet50 & \byteclass          & XGB & 0.7218 & 0.7313 & 0.7218 & 0.7223 \\
ViT      & \grayscale          & RF      & 0.7212 & 0.7286 & 0.7212 & 0.7196 \\
VGG16    & \byteclass          & MLP     & 0.7200 & 0.7312 & 0.7200 & 0.7211 \\
VGG16    & \hit                & RF      & 0.7182 & 0.7289 & 0.7182 & 0.7174 \\
ViT      & \byteclass          & XGB & 0.7176 & 0.7269 & 0.7176 & 0.7191 \\
ResNet50 & \hit                & RF      & 0.7171 & 0.7305 & 0.7171 & 0.7168 \\
VGG16    & \grayscale          & MLP     & 0.7153 & 0.7169 & 0.7153 & 0.7144 \\
ViT      & \grayscale          & XGB & 0.7147 & 0.7168 & 0.7147 & 0.7129 \\
ResNet50 & \hit                & XGB & 0.7141 & 0.7241 & 0.7141 & 0.7145 \\
VGG16    & \byteclass          & XGB & 0.7141 & 0.7260 & 0.7141 & 0.7167 \\
VGG16    & \byteclass          & LR      & 0.7129 & 0.7152 & 0.7129 & 0.7129 \\
ViT      & \hit                & RF      & 0.7118 & 0.7242 & 0.7118 & 0.7116 \\
VGG16    & \grayscale          & XGB & 0.7100 & 0.7167 & 0.7100 & 0.7097 \\
VGG16    & \hit                & XGB & 0.7100 & 0.7200 & 0.7100 & 0.7105 \\
VGG16    & \byteclassHilbert\  & RF      & 0.7076 & 0.7173 & 0.7076 & 0.7069 \\
ResNet50 & \grayscale          & MLP     & 0.7059 & 0.7135 & 0.7059 & 0.7074 \\
ResNet50 & \entropyHilbert            & RF      & 0.7059 & 0.7230 & 0.7059 & 0.7079 \\
ViT      & \entropyHilbert            & RF      & 0.7047 & 0.7138 & 0.7047 & 0.7048 \\
VGG16    & \byteclassHilbert\  & XGB & 0.7035 & 0.7140 & 0.7035 & 0.7054 \\
VGG16    & \grayscale          & LR      & 0.7029 & 0.7022 & 0.7029 & 0.7021 \\
VGG16    & \hit                & MLP     & 0.7024 & 0.7089 & 0.7024 & 0.6996 \\
ViT      & \byteclassHilbert\  & RF      & 0.7012 & 0.7075 & 0.7012 & 0.7004 \\
ResNet50 & \entropyHilbert            & XGB & 0.7006 & 0.7156 & 0.7006 & 0.7032 \\
VGG16    & \hit                & LR      & 0.6994 & 0.7020 & 0.6994 & 0.6996 \\
ViT      & \hit                & XGB & 0.6994 & 0.7084 & 0.6994 & 0.6996 \\
ResNet50 & \byteclass          & MLP     & 0.6988 & 0.7144 & 0.6988 & 0.7001 \\
VGG16    & \entropyHilbert            & RF      & 0.6988 & 0.7098 & 0.6988 & 0.7001 \\
VGG16    & \entropyHilbert            & XGB & 0.6988 & 0.7113 & 0.6988 & 0.7012 \\
ViT      & \grayscale          & MLP     & 0.6982 & 0.7009 & 0.6982 & 0.6975 \\
ResNet50 & \hit                & MLP     & 0.6976 & 0.7070 & 0.6976 & 0.6992 \\
ResNet50 & \byteclassHilbert\  & XGB & 0.6941 & 0.7078 & 0.6941 & 0.6961 \\
ResNet50 & \entropyHilbert            & MLP     & 0.6924 & 0.7055 & 0.6924 & 0.6937 \\
VGG16    & \entropyHilbert            & LR      & 0.6912 & 0.6979 & 0.6912 & 0.6929 \\
ViT      & \byteclassHilbert\  & XGB & 0.6900 & 0.6997 & 0.6900 & 0.6908 \\
ResNet50 & \byteclassHilbert\  & RF      & 0.6894 & 0.7056 & 0.6894 & 0.6905 \\
ViT      & \byteclass          & MLP     & 0.6888 & 0.6925 & 0.6888 & 0.6870 \\
ResNet50 & \hit                & LR      & 0.6865 & 0.6887 & 0.6865 & 0.6857 \\
ViT      & \entropyHilbert            & XGB & 0.6865 & 0.6976 & 0.6865 & 0.6867 \\
VGG16    & \byteclassHilbert\  & LR      & 0.6853 & 0.6885 & 0.6853 & 0.6860 \\
ResNet50 & \byteclassHilbert\  & MLP     & 0.6794 & 0.6944 & 0.6794 & 0.6785 \\
VGG16    & \entropyHilbert            & MLP     & 0.6788 & 0.6917 & 0.6788 & 0.6828 \\
VGG16    & \byteclassHilbert\  & MLP     & 0.6759 & 0.6830 & 0.6759 & 0.6747 \\
ResNet50 & \grayscale          & LR      & 0.6718 & 0.6737 & 0.6718 & 0.6697 \\
VGG16    & \grayscale          & KNN     & 0.6688 & 0.6677 & 0.6688 & 0.6662 \\
ViT      & \entropyHilbert            & MLP     & 0.6647 & 0.6694 & 0.6647 & 0.6639 \\
ViT      & \bigramCartesian  & MLP     & 0.6641 & 0.6723 & 0.6641 & 0.6652 \\
ResNet50 & \byteclass          & LR      & 0.6635 & 0.6666 & 0.6635 & 0.6624 \\
ViT      & \hit                & MLP     & 0.6624 & 0.6666 & 0.6624 & 0.6599 \\
ResNet50 & \grayscale          & KNN     & 0.6606 & 0.6592 & 0.6606 & 0.6588 \\
ViT      & \bigramCartesian  & RF      & 0.6571 & 0.6820 & 0.6571 & 0.6593 \\
ResNet50 & \byteclass          & KNN     & 0.6559 & 0.6526 & 0.6559 & 0.6491 \\
ViT      & \byteclass          & LR      & 0.6541 & 0.6601 & 0.6541 & 0.6538 \\
ResNet50 & \hit                & KNN     & 0.6518 & 0.6556 & 0.6518 & 0.6513 \\
VGG16    & \hit                & KNN     & 0.6488 & 0.6464 & 0.6488 & 0.6457 \\
VGG16    & \byteclass          & KNN     & 0.6488 & 0.6477 & 0.6488 & 0.6438 \\
ViT      & \bigramCartesian  & XGB & 0.6488 & 0.6645 & 0.6488 & 0.6502 \\
ViT      & \byteclassHilbert\  & MLP     & 0.6471 & 0.6546 & 0.6471 & 0.6467 \\
ViT      & \bigramCartesian  & KNN     & 0.6424 & 0.6457 & 0.6424 & 0.6417 \\
ViT      & \grayscale          & KNN     & 0.6382 & 0.6363 & 0.6382 & 0.6357 \\
ViT      & \byteclass          & KNN     & 0.6371 & 0.6342 & 0.6371 & 0.6321 \\
ResNet50 & \byteclassHilbert\  & LR      & 0.6353 & 0.6395 & 0.6353 & 0.6341 \\
ViT      & \grayscale          & LR      & 0.6347 & 0.6320 & 0.6347 & 0.6312 \\
ResNet50 & \entropyHilbert            & LR      & 0.6329 & 0.6364 & 0.6329 & 0.6315 \\
ViT      & \hit                & LR      & 0.6329 & 0.6344 & 0.6329 & 0.6318 \\
VGG16    & \byteclassHilbert\  & KNN     & 0.6324 & 0.6265 & 0.6324 & 0.6267 \\
VGG16    & \bigramCartesian  & MLP     & 0.6312 & 0.6318 & 0.6312 & 0.6269 \\
VGG16    & \entropyHilbert            & KNN     & 0.6306 & 0.6262 & 0.6306 & 0.6244 \\
ResNet50 & \entropyHilbert            & KNN     & 0.6300 & 0.6278 & 0.6300 & 0.6258 \\
ViT      & \byteclassHilbert\  & KNN     & 0.6300 & 0.6255 & 0.6300 & 0.6245 \\
ViT      & \hit                & KNN     & 0.6294 & 0.6285 & 0.6294 & 0.6273 \\
ViT      & \bigramPolar      & MLP     & 0.6276 & 0.6312 & 0.6276 & 0.6271 \\
VGG16    & \bigramCartesian  & KNN     & 0.6253 & 0.6300 & 0.6253 & 0.6237 \\
VGG16    & \bigramCartesian  & RF      & 0.6235 & 0.6480 & 0.6235 & 0.6252 \\
ResNet50 & \spiral             & RF      & 0.6218 & 0.6351 & 0.6218 & 0.6183 \\
ViT      & \entropyHilbert            & LR      & 0.6206 & 0.6173 & 0.6206 & 0.6173 \\
VGG16    & \bigramCartesian  & XGB & 0.6182 & 0.6328 & 0.6182 & 0.6191 \\
ViT      & \bigramPolar      & XGB & 0.6182 & 0.6267 & 0.6182 & 0.6175 \\
ResNet50 & \byteclassHilbert\  & KNN     & 0.6153 & 0.6114 & 0.6153 & 0.6101 \\
ViT      & \bigramPolar      & RF      & 0.6141 & 0.6394 & 0.6141 & 0.6163 \\
ViT      & \spiral             & RF      & 0.6129 & 0.6167 & 0.6129 & 0.6095 \\
ResNet50 & \bigramCartesian  & KNN     & 0.6124 & 0.6110 & 0.6124 & 0.6088 \\
VGG16    & \bigramCartesian  & LR      & 0.6112 & 0.6127 & 0.6112 & 0.6106 \\
ResNet50 & \bigramCartesian  & MLP     & 0.6094 & 0.6219 & 0.6094 & 0.6065 \\
ResNet50 & \spiral             & XGB & 0.6082 & 0.6170 & 0.6082 & 0.6058 \\
ViT      & \spiral             & XGB & 0.6059 & 0.6127 & 0.6059 & 0.6038 \\
ViT      & \bigramCartesian  & LR      & 0.6047 & 0.6068 & 0.6047 & 0.6048 \\
ViT      & \byteclassHilbert\  & LR      & 0.6000 & 0.5975 & 0.6000 & 0.5959 \\
ResNet50 & \bigramCartesian  & LR      & 0.5971 & 0.5968 & 0.5971 & 0.5940 \\
VGG16    & \spiral             & XGB & 0.5971 & 0.6049 & 0.5971 & 0.5955 \\
ResNet50 & \spiral             & KNN     & 0.5965 & 0.5989 & 0.5965 & 0.5942 \\
ResNet50 & \bigramCartesian  & XGB & 0.5941 & 0.6114 & 0.5941 & 0.5952 \\
VGG16    & \spiral             & RF      & 0.5935 & 0.6060 & 0.5935 & 0.5907 \\
ViT      & \bigramPolar      & KNN     & 0.5935 & 0.6051 & 0.5935 & 0.5950 \\
ResNet50 & \spiral             & MLP     & 0.5918 & 0.6165 & 0.5918 & 0.5925 \\
VGG16    & \spiral             & KNN     & 0.5912 & 0.5943 & 0.5912 & 0.5883 \\
ResNet50 & \bigramPolar      & KNN     & 0.5894 & 0.5900 & 0.5894 & 0.5875 \\
ResNet50 & \bigramPolar      & RF      & 0.5876 & 0.6078 & 0.5876 & 0.5851 \\
VGG16    & \grayscale          & SVM     & 0.5876 & 0.6043 & 0.5876 & 0.5813 \\
ResNet50 & \spiral             & LR      & 0.5859 & 0.5826 & 0.5859 & 0.5795 \\
ResNet50 & \bigramCartesian  & RF      & 0.5847 & 0.5997 & 0.5847 & 0.5807 \\
VGG16    & \hit                & SVM     & 0.5841 & 0.6047 & 0.5841 & 0.5818 \\
ResNet50 & \bigramPolar      & LR      & 0.5806 & 0.5822 & 0.5806 & 0.5789 \\
VGG16    & \spiral             & LR      & 0.5800 & 0.5742 & 0.5800 & 0.5732 \\
VGG16    & \bigramPolar      & MLP     & 0.5759 & 0.5819 & 0.5759 & 0.5758 \\
VGG16    & \spiral             & MLP     & 0.5741 & 0.5866 & 0.5741 & 0.5680 \\
VGG16    & \bigramPolar      & KNN     & 0.5729 & 0.5777 & 0.5729 & 0.5720 \\
ViT      & \bigramPolar      & LR      & 0.5729 & 0.5774 & 0.5729 & 0.5741 \\
ResNet50 & \bigramPolar      & XGB & 0.5706 & 0.5798 & 0.5706 & 0.5693 \\
ResNet50 & \bigramPolar      & MLP     & 0.5706 & 0.5923 & 0.5706 & 0.5749 \\
ViT      & \spiral             & MLP     & 0.5706 & 0.5692 & 0.5706 & 0.5626 \\
VGG16    & \bigramPolar      & RF      & 0.5676 & 0.5826 & 0.5676 & 0.5643 \\
VGG16    & \bigramPolar      & LR      & 0.5647 & 0.5707 & 0.5647 & 0.5658 \\
VGG16    & \byteclass          & SVM     & 0.5547 & 0.5695 & 0.5547 & 0.5499 \\
ViT      & \spiral             & KNN     & 0.5547 & 0.5509 & 0.5547 & 0.5501 \\
VGG16    & \bigramPolar      & XGB & 0.5535 & 0.5707 & 0.5535 & 0.5543 \\
VGG16    & \byteclassHilbert\  & SVM     & 0.5447 & 0.5525 & 0.5447 & 0.5340 \\
VGG16    & \bigramCartesian  & SVM     & 0.5394 & 0.5512 & 0.5394 & 0.5328 \\
ViT      & \grayscale          & SVM     & 0.5359 & 0.5424 & 0.5359 & 0.5252 \\
ViT      & \spiral             & LR      & 0.5324 & 0.5317 & 0.5324 & 0.5231 \\
ResNet50 & \hit                & SVM     & 0.5306 & 0.5541 & 0.5306 & 0.5253 \\
VGG16    & \entropyHilbert            & SVM     & 0.5294 & 0.5398 & 0.5294 & 0.5262 \\
ResNet50 & \bigramCartesian  & SVM     & 0.5041 & 0.5171 & 0.5041 & 0.4904 \\
ResNet50 & \grayscale          & SVM     & 0.4959 & 0.4944 & 0.4959 & 0.4729 \\
ViT      & \bigramCartesian  & SVM     & 0.4918 & 0.5026 & 0.4918 & 0.4784 \\
ViT      & \bigramPolar      & SVM     & 0.4900 & 0.4967 & 0.4900 & 0.4776 \\
ViT      & \byteclass          & SVM     & 0.4853 & 0.4840 & 0.4853 & 0.4710 \\
ViT      & \hit                & SVM     & 0.4829 & 0.5007 & 0.4829 & 0.4747 \\
VGG16    & \bigramPolar      & SVM     & 0.4824 & 0.4962 & 0.4824 & 0.4712 \\
ResNet50 & \byteclass          & SVM     & 0.4794 & 0.4756 & 0.4794 & 0.4634 \\
ResNet50 & \byteclassHilbert\  & SVM     & 0.4682 & 0.4793 & 0.4682 & 0.4584 \\
ResNet50 & \bigramPolar      & SVM     & 0.4665 & 0.4694 & 0.4665 & 0.4474 \\
VGG16    & \spiral             & SVM     & 0.4653 & 0.4696 & 0.4653 & 0.4408 \\
ResNet50 & \entropyHilbert            & SVM     & 0.4641 & 0.4595 & 0.4641 & 0.4487 \\
ResNet50 & \spiral             & SVM     & 0.4594 & 0.4564 & 0.4594 & 0.4314 \\
ViT      & \entropyHilbert            & SVM     & 0.4529 & 0.4601 & 0.4529 & 0.4318 \\
ViT      & \byteclassHilbert\  & SVM     & 0.4429 & 0.4875 & 0.4429 & 0.4271 \\
ViT      & \spiral             & SVM     & 0.3912 & 0.4123 & 0.3912 & 0.3578 \\
\hline\hline
\end{longtable}
}

\subsection{Full Custom CNN Results}\label{app:custom}

Table~\ref{tab:app_custom} contains the complete results for all~40 custom
CNN experiments across eight conversion types and five classifiers, sorted
by accuracy within each conversion type.

{\centering
\scriptsize
\begin{longtable}{lllcccc}
\caption{Complete custom CNN results on the test set (40 experiments)}\label{tab:app_custom} \\
\hline\hline
\textbf{Conversion} & \textbf{Classifier} &
\textbf{Accuracy} & \textbf{Precision} & \textbf{Recall} & \textbf{F1} \\
\hline
\endfirsthead
\hline
\textbf{Conversion} & \textbf{Classifier} &
\textbf{Accuracy} & \textbf{Precision} & \textbf{Recall} & \textbf{F1} \\
\hline
\endhead
\multirow{5}{*}{\grayscale}          & Random Forest       & 0.7518 & 0.7573 & 0.7518 & 0.7534 \\
& XGBoost  & 0.7500 & 0.7558 & 0.7500 & 0.7519 \\
& SVM      & 0.7459 & 0.7516 & 0.7459 & 0.7473 \\
& CatBoost & 0.7394 & 0.7448 & 0.7394 & 0.7411 \\
& MLP      & 0.7376 & 0.7450 & 0.7376 & 0.7391 \\
\hline
\multirow{5}{*}{\hit}                & MLP      & 0.7476 & 0.7568 & 0.7476 & 0.7493 \\
& SVM      & 0.7453 & 0.7526 & 0.7453 & 0.7468 \\
& CatBoost & 0.7453 & 0.7501 & 0.7453 & 0.7464 \\
& Random Forest        & 0.7441 & 0.7483 & 0.7441 & 0.7453 \\
& XGBoost  & 0.7341 & 0.7394 & 0.7341 & 0.7355 \\
\hline
\multirow{5}{*}{\entropyHilbert}           & Random Forest        & 0.7312 & 0.7353 & 0.7312 & 0.7315 \\
& XGBoost  & 0.7294 & 0.7335 & 0.7294 & 0.7299 \\
& SVM      & 0.7300 & 0.7356 & 0.7300 & 0.7306 \\
& CatBoost & 0.7300 & 0.7341 & 0.7300 & 0.7305 \\
& MLP      & 0.7235 & 0.7299 & 0.7235 & 0.7242 \\
\hline
\multirow{5}{*}{\byteclass}         & MLP      & 0.7176 & 0.7266 & 0.7176 & 0.7207 \\
& CatBoost & 0.7176 & 0.7258 & 0.7176 & 0.7199 \\
& Random Forest        & 0.7159 & 0.7220 & 0.7159 & 0.7172 \\
& SVM      & 0.7141 & 0.7243 & 0.7141 & 0.7172 \\
& XGBoost  & 0.7065 & 0.7145 & 0.7065 & 0.7084 \\
\hline
\multirow{5}{*}{\byteclassHilbert}  & CatBoost & 0.7100 & 0.7148 & 0.7100 & 0.7109 \\
& SVM      & 0.7094 & 0.7162 & 0.7094 & 0.7115 \\
& Random Forest        & 0.7071 & 0.7124 & 0.7071 & 0.7086 \\
& MLP      & 0.7035 & 0.7125 & 0.7035 & 0.7053 \\
& XGBoost  & 0.7024 & 0.7108 & 0.7024 & 0.7049 \\
\hline
\multirow{5}{*}{\bigramCartesian}  & CatBoost & 0.7200 & 0.7190 & 0.7200 & 0.7179 \\
& Random Forest        & 0.7206 & 0.7203 & 0.7206 & 0.7190 \\
& SVM      & 0.7188 & 0.7177 & 0.7188 & 0.7171 \\
& MLP      & 0.7188 & 0.7205 & 0.7188 & 0.7168 \\
& XGBoost  & 0.7106 & 0.7130 & 0.7106 & 0.7102 \\
\hline
\multirow{5}{*}{\bigramPolar}      & SVM      & 0.7159 & 0.7195 & 0.7159 & 0.7161 \\
& Random Forest        & 0.7118 & 0.7158 & 0.7118 & 0.7123 \\
& CatBoost & 0.7106 & 0.7124 & 0.7106 & 0.7102 \\
& MLP      & 0.7082 & 0.7187 & 0.7082 & 0.7083 \\
& XGBoost  & 0.6894 & 0.7063 & 0.6894 & 0.6928 \\
\hline
\multirow{5}{*}{\spiral}             & Random Forest        & 0.7035 & 0.7022 & 0.7035 & 0.7021 \\
& SVM      & 0.7029 & 0.7016 & 0.7029 & 0.7007 \\
& MLP      & 0.6994 & 0.6993 & 0.6994 & 0.6970 \\
& XGBoost  & 0.6947 & 0.6945 & 0.6947 & 0.6937 \\
& CatBoost & 0.6918 & 0.6934 & 0.6918 & 0.6912 \\
\hline\hline
\end{longtable}
}

\subsection{Ensemble Per-Class Classification Report}\label{app:ensemble_perclass}

Table~\ref{tab:app_perclass} contains the full per-class precision,
recall, and F1-score for the full 15-voter soft voting ensemble on
the test set. These results support the discussion in
Section~\ref{sec:perclass}.

\begin{table}[!htb]
\centering
\caption{Per-class classification report for the 15-voter ensemble (100 test samples per family)}\label{tab:app_perclass}
\begin{adjustbox}{scale=0.85}
\begin{tabular}{lcccr}
\toprule
\textbf{Family} & \textbf{Precision} & \textbf{Recall} & \textbf{F1} & \textbf{Support} \\
\midrule
\texttt{Agensla}  & 0.7174 & 0.6600 & 0.6875 & 100 \\
\texttt{Androm}  & 0.9189 & 0.6800 & 0.7816 & 100 \\
\texttt{Convagent}  & 0.8523 & 0.7500 & 0.7979 & 100 \\
\texttt{Crypt}  & 0.7976 & 0.6700 & 0.7283 & 100 \\
\texttt{Crysan}  & 0.9121 & 0.8300 & 0.8691 & 100 \\
\texttt{DCRat}  & 0.9900 & 0.9900 & 0.9900 & 100 \\
\texttt{Injuke}  & 0.7475 & 0.7400 & 0.7437 & 100 \\
\texttt{Makoob}  & 0.9340 & 0.9900 & 0.9612 & 100 \\
\texttt{Mokes}  & 0.7830 & 0.8300 & 0.8058 & 100 \\
\texttt{Noon}  & 0.8500 & 0.6800 & 0.7556 & 100 \\
\texttt{Remcos}  & 0.8602 & 0.8000 & 0.8290 & 100 \\
\texttt{Seraph}  & 0.5939 & 0.9800 & 0.7396 & 100 \\
\texttt{SnakeLogger}  & 0.9556 & 0.8600 & 0.9053 & 100 \\
\texttt{Stealerc}  & 0.6789 & 0.7400 & 0.7081 & 100 \\
\texttt{Strab}  & 0.8173 & 0.8500 & 0.8333 & 100 \\
\texttt{Taskun}  & 0.6752 & 0.7900 & 0.7281 & 100 \\
\texttt{Zenpak}  & 0.7745 & 0.7900 & 0.7822 & 100 \\
\midrule
Macro average & 0.8152 & 0.8018 & 0.8027 & 1{,}700 \\
\bottomrule
\end{tabular}
\end{adjustbox}
\end{table}


\phantomsection
\section*{Appendix C}\label{app:c}

\titleformat{\section}{\normalfont\large\bfseries}{}{0em}{#1\ \thesection}
\setcounter{section}{0}
\renewcommand{\thesection}{\Alph{section}}
\renewcommand{\thesubsection}{C.\arabic{subsection}}
\setcounter{table}{0}
\renewcommand{\thetable}{C.\arabic{table}}
\setcounter{figure}{0}
\renewcommand{\thefigure}{C.\arabic{figure}}

This appendix presents an additional ensemble experiment performed after
the defense, as recommended by the committee. It reports the complete ranked list of individual configurations across all three
tracks, together with precision, recall, and F1-score for each global pool,
extending the accuracy-only summary already given in Table~\ref{tab:ensemble_pools}
and Section~\ref{sec:voter_selection}.

Table~\ref{tab:global_ensemble_results} reports the soft-voting ensemble
performance for each pool. Table~\ref{tab:global_top25} lists the top~25 models 
in overall rank order.

\begin{table}[!htb]
\centering
\caption{Global soft voting pool results on the test set (1{,}700 samples, 17 families)}\label{tab:global_ensemble_results}
\begin{adjustbox}{scale=0.85}
\begin{tabular}{lccccc}
\toprule
\textbf{Pool} & \textbf{Voters} & \textbf{Accuracy} &
\textbf{Precision} & \textbf{Recall} & \textbf{F1} \\
\midrule
Global top-3   &  \zz3 & 0.7771 & 0.7885 & 0.7771 & 0.7780 \\
Global top-10  & 10 & 0.7806 & 0.7929 & 0.7806 & 0.7812 \\
Global top-15  & 15 & 0.7794 & 0.7841 & 0.7794 & 0.7790 \\
Global top-20  & 20 & 0.7782 & 0.7832 & 0.7782 & 0.7779 \\
Global top-25  & 25 & 0.7841 & 0.7889 & 0.7841 & 0.7835 \\
Global top-30  & 30 & 0.7853 & 0.7907 & 0.7853 & 0.7849 \\
\bottomrule
\end{tabular}
\end{adjustbox}
\end{table}

\begin{table}[!htb]
\centering
\caption{Top 25 models ranked by validation-selection accuracy across all tracks}\label{tab:global_top25}
\begin{adjustbox}{scale=0.85}
\begin{tabular}{clllc}
\toprule
\textbf{Rank} & \textbf{Track} & \textbf{Conversion} &
\textbf{Classifier} & \textbf{Accuracy} \\
\midrule
 \zz1 & Handcrafted     & \hit                         & Random Forest & 0.7559 \\
 \zz2 & Handcrafted     & \grayscale                   & Random Forest & 0.7535 \\
 \zz3 & Handcrafted     & \byteclassHilbert            & XGBoost       & 0.7488 \\
 \zz4 & Handcrafted     & \grayscale                   & XGBoost       & 0.7482 \\
 \zz5 & Handcrafted     & \hit                         & XGBoost       & 0.7471 \\
 \zz6 & Handcrafted     & \entropyHilbert              & Random Forest & 0.7441 \\
 \zz7 & Handcrafted     & \byteclassHilbert            & Random Forest & 0.7435 \\
 \zz8 & Handcrafted     & \byteclass                   & Random Forest & 0.7382 \\
 \zz9 & Handcrafted     & \byteclass                   & XGBoost       & 0.7306 \\
10 & Handcrafted     & \entropyHilbert              & XGBoost       & 0.7282 \\
11 & Custom CNN      & \grayscale                   & Random Forest & 0.7276 \\
12 & Custom CNN      & \hit                         & SVM           & 0.7241 \\
13 & Custom CNN      & \hit                         & XGBoost       & 0.7235 \\
14 & Custom CNN      & \hit                         & CatBoost      & 0.7235 \\
15 & Custom CNN      & \hit                         & Random Forest & 0.7235 \\
16 & Custom CNN      & \hit                         & MLP           & 0.7218 \\
17 & Custom CNN      & \grayscale                   & SVM           & 0.7218 \\
18 & Custom CNN      & \grayscale                   & CatBoost      & 0.7200 \\
19 & Custom CNN      & \grayscale                   & MLP           & 0.7200 \\
20 & Custom CNN      & \bigramCartesian             & SVM           & 0.7182 \\
21 & Custom CNN      & \grayscale                   & XGBoost       & 0.7176 \\
22 & Custom CNN      & \bigramCartesian             & CatBoost      & 0.7171 \\
23 & Pretrained CNN  & \grayscale\ (VGG16)          & Random Forest & 0.7171 \\
24 & Pretrained CNN  & \grayscale\ (ResNet50)       & Random Forest & 0.7165 \\
25 & Custom CNN      & \bigramCartesian             & Random Forest & 0.7165 \\
\bottomrule
\end{tabular}
\end{adjustbox}
\end{table}

As seen in Table~\ref{tab:global_ensemble_results}, the global pools all fall within
roughly one percentage point of one another, ranging from~77.7\%\ at 3 voters to a
peak of~78.5\%\ at 30 voters. As in the main experiments, global accuracy-based
ranking does not exceed the handcrafted-only pool, and adding further voters yields
only small gains: voters beyond the top handful are dominated by handcrafted and
custom-CNN models using conversions similar to those already present, reducing
diversity and limiting further gains, consistent with the diversity analysis in
Section~\ref{sec:diversity}.


\end{document}